\documentclass[%
 reprint,
 amsmath,amssymb,
 aps,
]{revtex4-2}

\usepackage{graphicx}

\usepackage{hyperref}

\usepackage{braket}
\usepackage{cleveref}
\Crefrangeformat{figure}{Fig.~#3#1#4 --~#5#2#6} 
\Crefrangeformat{table}{Tables~#3#1#4 --~#5#2#6} 
\Crefname{figure}{Fig.}{Fig.}
\Crefname{equation}{Eq.}{Eq.}
\creflabelformat{equation}{#2#1#3}
\usepackage{tikz}
\usetikzlibrary{matrix}
\usetikzlibrary{quantikz}
\usepackage{subfig}
\usepackage{todonotes}
\usepackage{ragged2e}
\usepackage{multirow}
\begin{document}

\title{Fighting noise with noise: a stochastic projective quantum eigensolver}
\author{Maria-Andreea Filip} \email{maf63@cam.ac.uk} 

\affiliation{ Yusuf Hamied Department of Chemistry, University of Cambridge,
Cambridge, UK }%
\date{\today}

\begin{abstract}

In the current noisy intermediate scale quantum  era of quantum computation, available hardware is severely limited by both qubit count and noise levels, precluding the application of many current hybrid quantum-classical algorithms to non-trivial quantum chemistry problems. In this paper we propose applying some of the fundamental ideas of conventional Quantum Monte Carlo algorithms --- stochastic sampling of both the wavefunction and the Hamiltonian --- to quantum algorithms in order to significantly decrease quantum resource costs. In the context of an imaginary-time propagation based projective quantum eigensolver, we present a novel approach to estimating physical observables which leads to a two order of magnitude reduction in the required sampling of the quantum state to converge the ground state energy of a system relative to current state-of-the-art eigensolvers. The method can be equally applied to excited-state calculations and, combined with stochastic approximations of the system Hamiltonian,  provides a promising near-term approach to  Hamiltonian simulation  for general chemistry on quantum devices.
\end{abstract}
\maketitle

\section{Introduction}
Electronic structure { problems are hard} for conventional computers, due to the exponential scaling of the Hilbert space with the physical size of the studied system. Quantum computers have been heralded as a solution to this problem,\cite{Feynman1982, Aspuru-Guzik2005} mainly due to their ability to use the quantum entanglement of their constituent qubits to encode even highly complex electronic wavefunctions in a linearly scaling qubit register. This promise has led to the development of a variety of quantum algorithms with applications in quantum chemistry, such as Hamiltonian simulation\cite{Berry2007, Berry2015, Low2017, Childs2018} and Quantum Phase Estimation.\cite{Abrams1997, Abrams1999}

However, while these techniques promise general solutions to many problems in quantum chemistry, they require many-qubit, fully fault-tolerant quantum computers to run. Currently we are in an era of Noisy Intermediate Scale Quantum (NISQ) hardware,\cite{Preskill2018} which is characterised by moderate numbers (10 -- 1000) of qubits prone to high error rates due to short coherence times and noisy gate implementations.\cite{Linke2017} While significant work is being carried out in the fields of quantum error correction\cite{Knill1197,Knill2000,Kribs2005,Terhal2015,Georgescu2020} and mitigation, \cite{Endo2021,Cai2022,Suzuki2022,Bultrini2023} the fully fault-tolerant quantum computer remains a thing of the future.

In order to make the most of currently available quantum resources, hybrid quantum-classical algorithms\cite{Farhi2014, Peruzzo2014, Colless2018, Higgott2019, Nakanishi2019, Greene-Diniz2021, Smart2021} have been developed, which shift most of the computational overhead onto a classical processor, while requiring specific, highly efficient operations to be carried out by the quantum processor. The most well known of these is the variational quantum eigensolver (VQE),\cite{Peruzzo2014} in which some parametrised wavefunction is encoded on a quantum device and used to compute a cost function --- usually the expectation value of a Hamiltonian operator. This is used in the classical optimisation of the parameters. In quantum chemistry, this approach has been used to calculate ground-state energies of molecules and model systems.\cite{Peruzzo2014, Shen2017, Hempel2018} 

{ Since methods like VQE require estimating expectation values of quantum observables, despite making use of only moderately sized quantum circuits they require a large number of repeated quantum measurements to obtain sufficiently accurate estimators. This is due to the intrinsic noise associated with the measurement of a quantum observable. Significant work has gone into devising more efficient measurement schemes,\cite{Hadfield2021,Choi2022, Choi2023, Yen2023} which obtain lower variances for the same number of measurements.}

In this paper, we approach the issue of decreasing quantum computational overhead of electronic structure problems by taking inspiration from conventional Quantum Monte Carlo (QMC) algorithms,\cite{Booth2009, Thom2010, Filip2020} which use noisy representations of the wavefunction and stochastic propagation techniques to obtain very high-accuracy results for challenging systems.  In particular, we consider a projective quantum eigensolver (PQE)\cite{Stair2021} approach and propose an algorithm to estimate the energy \textit{during} the propagation, which is found to require significantly fewer samples for the same accuracy as its conventional counterpart. 

{ In a perfectly noiseless regime, the PQE algorithm converges smoothly to the ground state of a given Ansatz. However, in practice, it processes data obtained from finitely many samples of a distribution stored in a quantum device, which is intrinsically stochastic even when ignoring NISQ-era device noise. Achieving a low enough uncertainty in this data to use it as if it were exact requires a large number of measurements. It is therefore compelling to employ instead an algorithm intended to operate on noisy measurements, such as QMC.} 

First, we establish a methodology to obtain accurate estimators for the ground state energy of a quantum system using relatively high-variance samples from an ongoing PQE propagation. We refer to this method as Monte Carlo PQE (MC-PQE). We then introduce additional stochastic approximations to further decrease quantum resource requirements. Stochastically rounding the wavefunction allows gates to be removed from the state preparation circuit, reducing its depth. The observables may also be estimated using a sampled subset of the Hamiltonian operator, lowering the number of independent measurements required. These techniques introduce additional noise into the algorithm, but we find this can be efficiently averaged out and, for a set of second row hydrides, show that the accuracy of the original algorithm can be recovered at significantly lower cost. 
Finally, we investigate the applicability of MC-PQE for excited states, using the folded-spectrum method, \cite{Peruzzo2014,Cao2019,CadiTazi2023} and find that, while convergence is somewhat more challenging, it can reduce the cost of an excited state calculation to that of the ground-state method. Given that the main drawback of the folded-spectrum method is its unfavourable scaling with system size, this is a very promising alternative.

In Sec. II we present the underlying theory of quantum eigensolvers, conventional QMC methods and the folded-spectrum approach which allows these algorithms to be expanded to excited state calculations. In Sec. III we describe the quantum algorithm employed in this paper, together with circuit implementations and possible stochastic variations. Finally, in Sec. IV we apply these methods to a range of molecular systems and discuss their performance. We draw our final conclusions in Sec. V.

\section{Background Theory}

\subsection{Variational Quantum Eigensolver}

One of the principal NISQ algorithms for quantum chemistry is the variational quantum eigensolver (VQE).\cite{Peruzzo2014} According to the variational principle, for a parametrised wavefunction $\Psi(\boldsymbol{\theta})$ that satisfies the boundary conditions of the problem, the expectation value of the energy is greater than or equal to the ground state energy of the system,
\begin{equation}
\braket{E(\boldsymbol{\theta})} = \frac{\braket{\Psi(\boldsymbol\theta)|\hat H | \Psi(\boldsymbol{\theta})}}{\braket{\Psi(\boldsymbol{\theta})| \Psi(\boldsymbol{\theta})}} \geq E_0.
\end{equation}
Therefore, by minimising $\braket{E(\boldsymbol{\theta})}$ with respect to the set of parameters $\boldsymbol{\theta}$, one can obtain the best possible approximation of a given form to the ground state wavefunction.
In VQE, the expectation value of the energy and any gradients used are computed by a quantum processor before being passed on to a classical optimisation algorithm.

There are various considerations to be taken into account when choosing parametrised wavefunction Ans\"atze for VQE. Heuristic hardware efficient Ans\"atze (HEA)\cite{Kandala2017} attempt to introduce many degrees of freedom while maintaining low circuit depths. However, optimisation on the resulting landscapes is often difficult, as they are plagued by barren plateaus\cite{McClean2018, Fontana2023} --- regions of parameter space in which the gradient vanishes exponentially with system size -- that can prevent the optimisation algorithm from converging to a true minimum. { The propensity for a problem to display barren plateaus is dependent on a variety of system and algorithm properties\cite{Ragone2023} and} has been shown to be worse for more expressive Ans\"atze \cite{Holmes2022} and higher degrees of entanglement.\cite{Marrero2021} However, the intrinsic noise resulting from finite sampling of the quantum state can be helpful in improving optimisation outcomes.\cite{Liu2023} 

Alternatively, one can use physically motivated Ans\"atze such as unitary coupled cluster (UCC),\cite{Kutzelnigg1982,Kutzelnigg1983,Kutzelnigg1984,Bartlett1989} in which the wavefunction is expressed as
\begin{equation}
\ket{\Psi} = e^{\hat \tau}\ket{\phi_0} = e^{\hat T - \hat T^\dagger}\ket{\phi_0},
\label{eq:UCC}
\end{equation}
where { $\ket{\phi_0}$ is typically the Hartree--Fock wavefunction,}
\begin{align}
\hat T = \sum_\mathbf{i} t_\mathbf{i} \hat a_\mathbf{i},\\
\hat T^\dagger = \sum_\mathbf{i} t_\mathbf{i} \hat a_\mathbf{i}^\dagger,
\end{align}
{$\mathbf{i}$ is an index running over all determinants in the Hilbert space and $\hat a_\mathbf{i}$ and $\hat a_\mathbf{i}^\dagger$ are  fermionic excitation and de-excitation operators respectively, such that ${\hat a_\mathbf{i}\ket{\phi_0} = \pm\ket{\phi_\mathbf{i}}}$ and ${\hat a_\mathbf{i}^\dagger\ket{\phi_i} = \pm\ket{\phi_0}}$, with signs derived from the anti-commutation relations of creation and annihilation operators. The cluster amplitudes $t_\mathbf{i}$ are optimised to give an approximation of the ground state wavefunction. In principle, this wavefunction form can include up to all-electron (de)excitations, but it is common to truncate it to some low excitation order, most commonly including only up to two-electron operators, which gives the unitary coupled cluster singles and doubles (UCCSD) Ansatz.} This type of Ansatz is generally found to be easier to optimise than HEAs but has its own challenges. In order to express the UCC operator in the set of gates available on most quantum computers, one generally has to perform a Trotter decomposition\cite{Trotter1959, Suzuki1976} of the cluster operator,
\begin{equation}
 e^{\hat \tau} = \lim_{\rho \rightarrow \infty}\Big(\prod_\mathbf{i} e^{\hat\tau_\mathbf{i}/\rho}\Big)^\rho,
\end{equation}
{ where $\hat \tau_\mathbf{i} = t_\mathbf{i}(a_\mathbf{i} - a_\mathbf{i}^\dagger)$ and we have rewritten the exponential in \Cref{eq:UCC} as ${\mathrm{exp}(\hat T - \hat T^\dagger) = \mathrm{exp}(\sum_\mathbf{i} t_\mathbf{i}(a_\mathbf{i} - a_\mathbf{i}^\dagger)) = \mathrm{exp}(\sum_\mathbf{i}\hat\tau_\mathbf{i})}$.}
Usually, rather than working in the large $\rho$ limit, $\rho$ is taken to be unity, leading to a disentangled { form} of the UCC Ansatz, which has been found to be as expressive as the original algorithm in most circumstances.\cite{Evangelista2019}

Because of this decomposition, UCC-based circuits often tend to be prohibitively deep for NISQ devices. { Various alternative algorithms have been devised, that attempt to enforce some of the known physical properties of the wavefunction to generate easier to optimise energy functions, while maintaining a reduced circuit depth relative to UCC. Examples include} symmetry preserving Ans\"atze,\cite{Gard2020} adaptive Ans\"atze\cite{Grimsley2019b} and methods like qubit coupled cluster (QCC)\cite{Ryabinkin2018,Ryabinkin2020} and Givens-rotation-based Ans\"atze\cite{Arrazola2022} which avoid some of the complexities of fermionic excitation operators.
\subsection{Projective Quantum Eigensolver}
As an alternative to variational algorithms, one can solve electronic structure problems using projective approaches. In classical settings, this technique is most commonly employed in Projector Monte Carlo (PMC) algorithms, which will be discussed in more detail in \cref{sec:qmc}.

Consider a trial wavefunction expressed as \mbox{$\ket{\Psi_0} =\hat U (\boldsymbol\theta)\ket{\phi_0}$}, where $\ket{\phi_0}$ is a simple reference wavefunction such as a Hartree--Fock determinant and $\hat U(\mathbf{\boldsymbol\theta})$ is some parametrised unitary operator.
If this were an exact solution to the Schr\"odinger equation, then
\begin{equation}
\hat H \hat U \ket{\phi_0} = E \hat U\ket{\phi_0}
\end{equation}
and
\begin{equation}
\hat U^\dagger \hat H \hat U \ket{\phi_0} = E \ket{\phi_0}.
\label{eq:linked}
\end{equation}
Projecting \Cref{eq:linked} onto the reference wavefunction $\ket{\phi_0}$ gives
\begin{equation}
    \braket{\phi_0|\hat U^\dagger \hat H \hat U |\phi_0} = E_\mathrm{PQE},
    \label{eq:epqe}
\end{equation}
while projecting onto the complete set of $\ket{\phi_\mathbf{i}}$ orthogonal to $\ket{\phi_0}$ gives a set of { quantities referred to as residuals}
\begin{equation}
 r^\mathrm{l}_\mathbf{i} = \braket{\phi_\mathbf{i}|\hat U^\dagger \hat H \hat U |\phi_0}.
 \label{eq:lk_resid}
\end{equation}
{If  $\hat U\ket{\phi_0}$ is an exact eigenfunction of $\hat H$, then substituting \Cref{eq:linked} into \Cref{eq:lk_resid} gives
\begin{equation}
    r^\mathrm{l}_\mathbf{i} = E\braket{\phi_\mathbf{i}|\phi_0} = 0
     \label{eq:resid}
\end{equation}
for any $\phi_\mathbf{i}$ in the Hilbert space. Therefore, the idea behind projective algorithms is to use the condition given by \Cref{eq:resid} to obtain a set of equations that can be solved for the values of the parameters in $\hat U(\boldsymbol \theta)$, in order to obtain an approximation to the true ground state. This is done by enforcing \Cref{eq:resid} for a subset of states $\ket{\phi_\mathbf{i}}$, corresponding to the parameters in $\hat U(\boldsymbol \theta)$.}  Solving \Cref{eq:epqe,eq:resid} for this subset leads to a Projective Quantum Eigensolver (PQE)\cite{Stair2021} algorithm.

If $\hat U$ is a UCC cluster operator and $\ket{\phi_0}$ is the Hartree--Fock determinant, \Cref{eq:epqe,eq:resid} can be solved using a quasi-Newton iterative scheme such as\cite{Stair2021} 
\begin{equation}
\theta_\mathbf{i}^{(n+1)} =\theta_\mathbf{i}^{(n)} + \frac{r^\mathrm{l}_\mathbf{i}}{\Delta_\mathbf{i}}
\end{equation}
{ If $\ket{\phi_{\mathbf i}}$ labels $\ket{\phi_{ij...}^{ab...}}$, with indices $i,j, ...$ ( $a, b, ...$) denoting orbitals occupied (unoccupied) in the HF determinant, from (to) which electrons are excited, then  ${\Delta_\mathbf{i}} = \epsilon_i + \epsilon_j + ... - \epsilon_a - \epsilon_b - ...$, where $\epsilon_i$ are Hartree--Fock orbital energies and the indices correspond to the same orbitals as above}. This algorithm is appealing for NISQ device use because, unlike { energy gradients calculated in many VQE algorithms,} the residuals $r^\mathrm{l}_\mathbf{i}$ can be computed as expectation values at the same cost as the overall energy, since
\begin{align}
\begin{split}
r^\mathrm{l}_\mathbf{i} = &\braket{\Omega_\mathbf{i}(\pi/4)|\hat U^\dagger \hat H \hat U|\Omega_\mathbf{i}(\pi/4)} \\
&- \frac{1}{2}\braket{\phi_0|\hat U^\dagger \hat H \hat|\hat U \phi_0} -\frac{1}{2}\braket{\phi_\mathbf{i}|\hat U^\dagger \hat H \hat U|\phi_\mathbf{i}}
\end{split}
\label{eq:exp_resid}
\end{align}
where $\ket{\Omega_\mathbf{i}(\theta)} = e^{\theta \hat \tau_\mathbf{i}}\ket{\phi_0} = e^{\theta (\hat a_\mathbf{i} - \hat a_\mathbf{i}^\dagger)}\ket{\phi_0} = {\cos{\theta}\ket{\phi_0} + \sin{\theta}\ket{\phi_\mathbf{i}}}$.

\subsection{Quantum Monte Carlo}\label{sec:qmc}

As mentioned above, the main use of projective methods in conventional quantum chemistry is in PMC algorithms, which are based on the imaginary-time Schr\"odinger equation,
\begin{equation}
    \frac{\partial \ket{\Psi}}{\partial \beta} = - \hat H \ket{\Psi},
    \label{eq:im_sch}
\end{equation}
where $\beta$ is imaginary time and $\hat H$ is the Hamiltonian of the system of interest. The lowest-energy solution to \Cref{eq:im_sch}, $\ket{\Psi_0}$, is given by
\begin{equation}
    \ket{\Psi_0} = \lim_{\beta \rightarrow \infty} e^{-\beta (\hat H - E_0)}\ket{\Psi(0)},
\end{equation}
where $\ket{\Psi(0)}$ is some initial trial wavefunction such that $\braket{\Psi_0|\Psi(0)}\neq0$ and $E_0$ is the lowest eigenvalue of $\hat H$. The time-evolution operator can also undergo a Trotter expansion, giving
\begin{equation}
    \ket{\Psi_0} = \lim_{n \rightarrow \infty} (e^{-\Delta \beta (\hat H - E_0)})^n \ket{\Psi(0)},
\end{equation}
where $\Delta \beta = \beta/n$. For small $\Delta \beta$, the exponential can be approximated by a linear expansion, leading to the form of the imaginary time propagator most commonly used in projector methods,
\begin{equation}
    \ket{\Psi_0} = \lim_{n \rightarrow \infty} (1 - \Delta\beta (\hat H - E_0))^n \ket{\Psi(0)}.
    \label{eq:proj}
\end{equation}
Considering a single time-step $\Delta \beta$, one can write
\begin{equation}
        \ket{\Psi(\beta + \Delta \beta} = (1 - \Delta\beta (\hat H - E_0)) \ket{\Psi(\beta)},
\end{equation}
which describes the step-wise imaginary time evolution of the trial wavefunction $\ket{\Psi}$. As in PQE, this equation can be projected onto the various Slater determinants in the Hilbert space to give
\begin{equation}
    \braket{\phi_\mathbf{i}|\Psi(\beta + \Delta \beta)} = \braket{\phi_\mathbf{i}|1 - \Delta\beta (\hat H - E_0)|\Psi(\beta)}
    \label{eq:pmc}
\end{equation}

{ In the original Full Configuration Interaction Quantum Monte Carlo (FCIQMC) algorithm,\cite{Booth2009} the wavefunction is parametrised as a linear expansion in all determinants in the Hilbert space, $\ket{\Psi} = \sum_\mathbf{i}c_\mathbf{i}\ket{\phi_\mathbf{i}}$. \Cref{eq:pmc} can then be approximated as
\begin{equation}
    c_\mathbf{i}(\beta + \Delta \beta) = c_\mathbf{i}(\beta) - \Delta\beta\braket{\phi_\mathbf{i}|\hat H-E_0|\Psi(\beta)}
\end{equation}
or
\begin{equation}
  c_\mathbf{i}(\beta + \Delta \beta) = c_\mathbf{i}(\beta) - \Delta\beta\Big(c_\mathbf{i} (H_\mathbf{ii} - E_0) + \sum_{\mathbf{j} \neq \mathbf{i}} c_\mathbf{j}H_\mathbf{ij}\Big),  
  \label{eq:pop_dyn}
\end{equation}
where $H_\mathbf{ij} = \braket{\phi_\mathbf{i}|\hat H|\phi_\mathbf{j}}$. We note that this is equivalent to 
\begin{equation}
c_\mathbf{i}(\beta + \Delta \beta) = c_\mathbf{i}(\beta) - \Delta\beta r^\mathrm{u}_\mathbf{i},
\label{eq:param_resid}
\end{equation}
where { 
${r^\mathrm{u}_\mathbf{i} = \braket{\phi_\mathbf{i}|\hat H-E_0|\Psi(\beta)}}$.} 

In FCIQMC, \Cref{eq:pop_dyn} is interpreted as governing the population dynamics of a set of particles (``walkers", ``psips" or ``excips") living in the Hilbert space, such that on each determinant there exists a particle population proportional to its configuration interaction (CI) coefficient, $c_\mathbf{i}$. Rather than iterating these equations exactly, their terms are sampled stochastically and can be described by three processes:\cite{Booth2009}
\begin{itemize}
\item \textbf{Spawning} of particles from determinant $\ket{\phi_\mathbf{j}}$ to  $\ket{\phi_\mathbf{i}}$ corresponds to the off-diagonal action of the Hamiltonian.
\begin{equation} 
\delta c_\mathbf{i}^{(s, k)} = - \Delta\beta \frac{c_\mathbf{j}H_\mathbf{ij}}{p_\mathrm{select}(\phi_\mathbf{j}) p_\mathrm{gen}(\mathbf{i}|\mathbf{j}) n_\mathrm{attempts}},
\label{eq:pspawn}
\end{equation}
where $p_\mathrm{select}(\phi_\mathbf{j})$ is the probability of selecting $\phi_\mathbf{j}$ as the source of the spawn, $p_\mathrm{gen}(\mathbf{i}|\mathbf{j})$ is the conditional probability of selecting $\phi_\mathbf{i}$ as the target of a spawn given it originates from $\phi_\mathbf{j}$ and $n_\mathrm{attempts}$ is the number of spawning attempts. $k$ is the index of the current attempt and $s$ denotes that this is a spawning event.
\item \textbf{Death} or cloning from $\ket{\phi_\mathbf{i}}$ to itself corresponds to the diagonal action of the Hamiltonian.
\begin{equation} 
\delta c_\mathbf{i}^{(d, k)} = - \Delta\beta \frac{c_\mathbf{i} (H_\mathbf{ii} - S)}{p_\mathrm{select}(\phi_\mathbf{i}) n_\mathrm{attempts}},
\end{equation}
where $d$ denotes that this is now a death event.
\item \textbf{Annihilation} ensures that the two equivalent oppositely signed versions of the wavefunction do not proliferate simultaneously and corresponds to collecting all contributions to $c_\mathbf{i}$ into a single value.
\begin{equation}
c_\mathbf{i} = c_\mathbf{i} + \sum_k^{n_\mathrm{attempts}}\delta c_\mathbf{i}^{(s, k)} + \sum_k^{n_\mathrm{attempts}}\delta c_\mathbf{i}^{(d, k)}
\end{equation}
\end{itemize}}

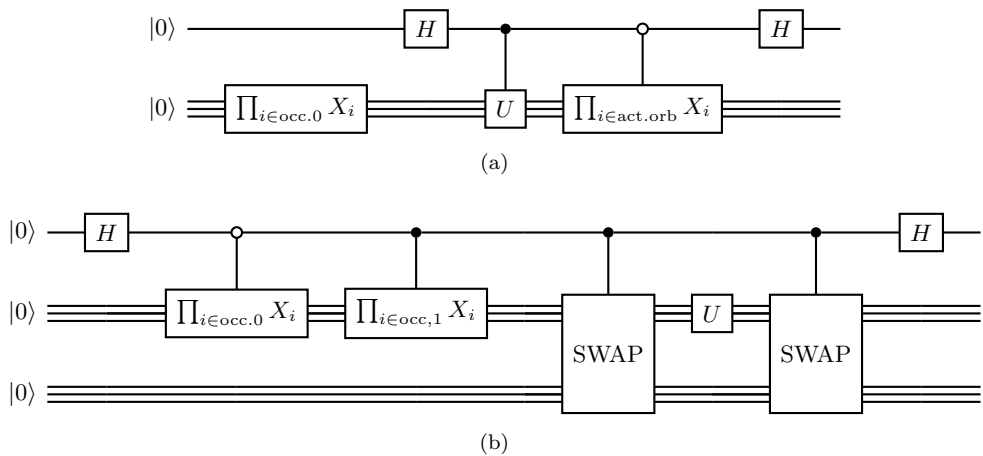
\begin{figure*}
    \centering
\subfloat[][ ]{\begin{quantikz}
\lstick{$\ket{0}$} & \qw& \gate{H}& \ctrl{1}  &\octrl{1}&\gate{H}&\qw\\
\lstick{$\ket{0}$}&\gate{\prod_{i \in \mathrm{occ. 0}} X_i}\qwbundle[alternate]{}&\qwbundle[alternate]{} &\gate{U}\qwbundle[alternate]{}&\gate{\prod_{i \in \mathrm{act. orb}} X_i} \qwbundle[alternate]{}&\qwbundle[alternate]{}&\qwbundle[alternate]{}
\end{quantikz}}\\
\subfloat[][ ]{\begin{quantikz}
\lstick{$\ket{0}$} &  \gate{H}& \octrl{1}  &\ctrl{1}&\qw&\ctrl{1}&\qw&\ctrl{1}&\gate{H}&\qw\\
\lstick{$\ket{0}$}&\qwbundle[alternate]{} &\gate{\prod_{i \in \mathrm{occ. 0}} X_i}\qwbundle[alternate]{}&\gate{\prod_{i \in \mathrm{occ, 1}} X_i} \qwbundle[alternate]{}&\qwbundle[alternate]{}&\gate[2]{\mathrm{SWAP}}\qwbundle[alternate]{}&\gate{U}\qwbundle[alternate]{}&\gate[2]{\mathrm{SWAP}}\qwbundle[alternate]{}&\qwbundle[alternate]{}&\qwbundle[alternate]{}\\
\lstick{$\ket{0}$}&\qwbundle[alternate]{}&\qwbundle[alternate]{}&\qwbundle[alternate]{}&\qwbundle[alternate]{}&\qwbundle[alternate]{}&\qwbundle[alternate]{}&\qwbundle[alternate]{}&\qwbundle[alternate]{}&\qwbundle[alternate]{}
\end{quantikz}}
    \caption{\small \justifying Quantum circuits for measuring the overlap $s_\mathrm{i} = \braket{\phi_\mathbf{i}|\hat U \phi_0}$. The first is a modified Hadamard test,\cite{quantumalgorithms} which uses a single ancilla qubit and a controlled version of the unitary $\hat U$. The second was proposed by Huggins \textit{et al.}\cite{Huggins2020} and needs $(n_\mathrm{qubits} + 1)$ ancilla but no controlled $\hat U$.}
    \label{fig:circs}
\end{figure*}

We note that in the death step the unknown true ground state energy $E_0$ has been replaced by the shift $S$. This acts as a population control parameter and in standard QMC procedures eventually converges to a stochastic estimator of $E_0$.\cite{Booth2009} The projected energy
\begin{equation}
    E_\mathrm{proj} = \frac{\braket{\phi_0|\hat H| \Psi(\beta)}}{\braket{\phi_0|\Psi(\beta)}}.
    \label{eq:e_proj}
\end{equation}
is normally used as a second way to estimate $E_0$.

{ By employing a stochastic representation of the wavefunction, algorithms like FCIQMC take advantage of the sparsity of the Hamiltonian, and are therefore able to tackle problems far beyond those tractable for a conventional FCI solvers.\cite{Weser2022} This behaviour can be further enhanced by developments such as the initiator approximation\cite{Cleland2010} and the adaptive-shift method,\cite{Ghanem2019, Ghanem2020} which artificially stochastically increase the sparsity of the Hamiltonian.}

{Alternative projective Monte Carlo formulations have been developed using Coupled Cluster (CC)\cite{Thom2010} and UCC\cite{Filip2020} parametrisations of the wavefunction. These are based on the fact that for an exponential Ansatz, if a determinant $\ket{\phi_\mathbf{i}}$ is reachable by an excitation included in the cluster operator, then $\braket{\phi_\mathbf{i}|\Psi} = t_\mathbf{i} + \mathcal{O}(t^2)$, where $t_\mathbf{i}$ are the cluster amplitudes. Since these methods are only valid in the regime in which the cluster amplitudes are small, a similar equation to \Cref{eq:pop_dyn} can be written down as
\begin{equation}
      t_\mathbf{i}(\beta + \Delta \beta) = t_\mathbf{i}(\beta) - \Delta\beta\Big(c_\mathbf{i} (H_\mathbf{ii} - E_0) + \sum_{\mathbf{j} \neq \mathbf{i}} c_\mathbf{j}H_\mathbf{ij}\Big).  
  \label{eq:pop_dyn2}
\end{equation}
Therefore, similar stochastic propagation techinques can be used as in the case of FCIQMC, however one only has access to the cluster amplitudes $t_\mathbf{i}$. The corresponding CI coefficients must also be sampled stochastically, leading to additional complexity in the selection routine, which is responsible for a significant portion of the computational cost of these algorithms.}

In FCIQMC it is easy to cycle through all determinants present in the wavefunction or sample them uniformly. In contrast, in Coupled Cluster Monte Carlo (CCMC) a doubly excited determinant $\ket{\phi_\mathbf{i}} = \ket{\phi_{ij}^{ab}}$ will have a CI coefficient given by
\begin{equation}
    c_{ij}^{ab} = t_{ij}^{ab} + t_{i}^at_j^b - t_i^bt_j^a,
\end{equation}
where $t_{i}^a$, $t_{ij}^{ab}$ are the amplitudes of the corresponding single and double excitation operators $\hat a_i^a$ and $\hat a_{ij}^{ab}$ in the cluster operator $\hat T$. Therefore, when selecting a determinant for spawning or death, one must consider both composite (e.g. $t_{i}^at_j^b$) and non-composite ($t_{ij}^{ab}$) contributions (``clusters") to it. The standard way to do this in CCMC\cite{Thom2010, Scott2017} is to:
\begin{enumerate}
    \item Select a cluster size $s$ with some probability $p_s$;
    \item (Optionally) select a particular type of cluster of size $s$ by imposing constraints on the excitation levels of the individual excitors involved;\cite{Scott2017}
    \item Select $s$ excitors $e$ to form the cluster, each with probability $p_e \propto t_e$;
    \item ``Collapse" the excitors to determine the corresponding determinant (e.g. $\hat a_i^a \hat a_j^b \ket{\phi_0} = \pm\ket{\phi_{ij}^{ab}}$).
\end{enumerate}

A variety of selection algorithms have been developed for CCMC  to better importance sample the underlying wavefunction.\cite{Scott2017} Nevertheless, this remains one of its main challenges, particularly when expanded to high truncation levels\cite{Neufeld2017} or multi-reference formulations.\cite{Filip2019} 

To further complicate matters, in unitary CCMC (UCCMC)\cite{Filip2020} the maximum cluster size is formally infinite, due to alternating excitation and de-excitation operators in the clusters. While results converge relatively quickly with maximum cluster size, the inclusion of large composite clusters leads to instabilities in the Monte Carlo propagation.

\subsection{Folded-spectrum} 
Methods like VQE, PQE and standard PMC are all intended to find the ground state of a system of interest. The energy of some excited states can be found with these techniques by enforcing orthogonality between the state of interest and the ground state, for example by considering different symmetry sectors of the Hilbert space.\cite{Nakanishi2019} However, obtaining general excited states is not-trivial.

{ Using the FCIQMC approach, it is possible to obtain multiple excited states simultaneously by instantaneously orthogonalising the stochastic wavefunction representations of multiple independent FCIQMC calculations.\cite{Blunt2015} A similar idea forms the basis for the Variational Quantum Deflation (VQD)\cite{Higgott2019} algorithm, although this requires sequential optimisation of the different excited states. Alternative hybrid excited state algorithms include the quantum subsbace expansion (QSE) method,\cite{McClean2017,Colless2018, Takeshita2020} in which the Hamiltonian is diagonalised in some subset of the Hilbert space to obtain estimates of low-lying eigenstates; the multistate-contracted VQE method,\cite{Parrish2019} which optimises an entanglement matrix between some approximate excited states obtained from a conventional quantum chemistry calculation; or the witness-assisted variational eigenspectra solver (WAVES),\cite{Santagati2018} which uses the von Neumann entropy of states to ascertain whether they are eigenstates of the Hamiltonian.}

Another approach to directly target excited states is the folded-spectrum method,\cite{Peruzzo2014,Cao2019,CadiTazi2023} in which the system Hamiltonian $\hat H$ is replaced by
\begin{equation}
\hat H_\mathrm{fs} (\omega) =  (\hat H - \omega)^2,
\label{eq:fs_ham}
\end{equation}
which has the same eigenstates as the original $\hat H$. However, the eigenvalues become $(E_i - \omega)^2$ so the ground state of $\hat H_\mathrm{fs}(\omega)$ becomes the eigenstate with true energy $E_i$ closest to $\omega$. Therefore, by changing $\omega$ one can target any stationary state of the system with conventional variational or projective techniques. { This method has been previously used in conjunction with FCIQMC\cite{Booth2012}} and similar approaches have been used in the conventional Variational Monte Carlo (VMC) community under the name of state-specific variance optimisation.\cite{Umrigar1988, Zhao2016,Shea2017,Cuzzocrea2020, Otis2023}

In variational folded-spectrum approaches, one minimises the quantity
\begin{equation}
\braket{E}_\mathrm{fs} (\omega) = \frac{\braket{\Psi(\boldsymbol{\theta})|(\hat H - \omega)^2| \Psi(\boldsymbol{\theta})}}{\braket{\Psi(\boldsymbol{\theta})|\Psi(\boldsymbol{\theta})}},
\end{equation}
while in projective approaches one updates the parameters using the residuals
\begin{equation}
r^\mathrm{fs}_\mathbf{i} = \braket{\phi_\mathbf{i}|(\hat H - \omega)^2 -E_\mathrm{fs,0}| \Psi(\boldsymbol{\theta})}.
\label{eq:res_fs}
\end{equation}
The algorithms employed are equivalent to the standard ground-state approaches, but the folded-spectrum Hamiltonian is more complex than the original Hamiltonian. { Consider a standard chemistry Hamiltonian
\begin{equation}
    \hat H = \sum_{pq} f_{pq} \hat p^\dagger \hat q + \sum_{pqrs}g_{pqrs}\hat p^\dagger \hat q ^\dagger \hat r \hat s,
    \label{eq:ferm_hamil}
\end{equation}
where $\hat p$ and $\hat p^\dagger$ are fermionic creation and annihilation operators respectively for spin-orbital $p$ and $f_{pq}$ and $g_{pqrs}$ are the one- and two-body contributions to the Hamiltonian. The} corresponding folded-spectrum Hamiltonian will include up to four-body terms, 
\begin{equation}
    \sum_{pqrstuvw}g_{pqrs}g_{tuvw}\hat p^\dagger \hat q ^\dagger \hat t^\dagger \hat s^\dagger \hat r \hat s \hat v \hat w,
\end{equation}
which will be significantly more expensive to compute than the one- and two-body terms of the original Hamiltonian.{ The total number of terms in the operator will therefore scale as $N^8$ in the number of one-electron basis functions, rather than the $N^4$ scaling of the original Hamiltonian. Due to this significant cost increase, the folded-spectrum technique is currently limited to only small systems.}
\begin{figure*}
\centering
\begin{quantikz}
\lstick{$|a\rangle$} & \qw & \qw & \qw  &\ctrl{4} & \qw & \qw &\qw & \qw \\
\lstick{$|q_0\rangle$} & \qw & \qw & \qw  &\qw & \qw & \qw  & \qw& \qw \\
\lstick{$|q_1\rangle$} & \gate{H} & \ctrl{1}  & \qw & \qw & \qw & \ctrl{1}   & \gate{H} & \qw \\
\lstick{$|q_2\rangle$} & \qw & \targ{} & \ctrl{1}   & \qw & \ctrl{1} & \targ{} & \qw  & \qw \\
\lstick{$|q_3\rangle$} & \gate{R_x(\frac{3\pi}{2})} &\qw & \targ{}  & \gate{R_z(\theta)} & \targ{} & \qw & \gate{R_x(\frac{\pi}{2})} & \qw \\
\end{quantikz}
\caption{\small Controlled Pauli gadget encoding $e^{-i\frac{\theta}{2}( X_1 \otimes Z_2\otimes Y_3)}$.}
\label{fig:controlled_pauli}
\end{figure*}
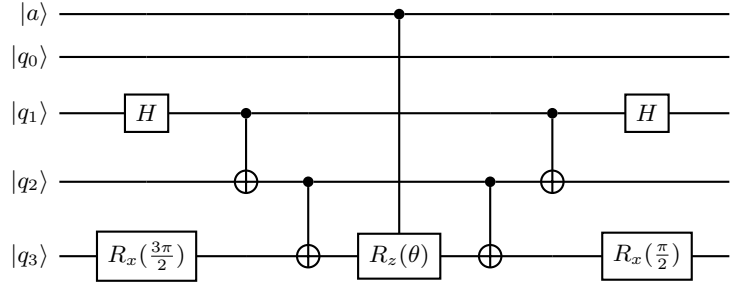
\section{Quantum Algorithms}
\subsection{Unlinked Residual Measurement Implementation}\label{sec:algorithm}
We begin by discussing the implementation of quantum circuits to compute the unlinked residual 
\begin{equation}
    r^\mathrm{u}_\mathbf{i} = \braket{\phi_\mathbf{i}|\hat H-S|\Psi(\beta)}
\end{equation}
and the overlap
\begin{equation}
    s_\mathrm{i} = \braket{\phi_\mathbf{i}|\Psi(\beta)}
\end{equation}
required by the QMC methods discussed in Sec. \ref{sec:qmc}.

As detailed in Sec. \ref{sec:qmc}, exponential Ansatz-based QMC algorithms require complex selection algorithms to estimate $r_\mathbf{i}^\mathrm{u}$, as computing it directly would require storing the full CI wavefunction, which is not memory-efficient. For large cluster sizes, this often becomes the limiting factor on the speed and rate of convergence of a calculation. Therefore, using a quantum device to compute these terms without the memory overhead would be a significant boon to these methods. This has been shown in the context of auxilliary-field QMC (AFQMC)\cite{Zhang2003}, for which overlaps with complex trial wavefunctions can be efficiently calculated using a quantum computer.\cite{Huggins2022} 

The standard implementation of the UCC Ansatz in the quantum circuit representation is to map Fermionic creation and annihilation operators onto strings of Pauli $X$, $Y$ and $Z$ matrices (or gates). There are multiple standard mapping schemes,\cite{Jordan1928,Bravyi2002,Seeley2012,Tranter2015} of which we employ here the Jordan--Wigner approach,\cite{Jordan1928} in which

\begin{equation}
\begin{split}
\hat p^\dagger = \sigma_p^+ \bigotimes_{q>p} Z_q \\
\hat p = \sigma_p^- \bigotimes_{q>p} Z_q,
\end{split}
\end{equation}
where the product of $Z$ matrices encodes the required parity to preserve fermionic anti-commutation relations and $\sigma_p^\pm = X_p \pm iY_p$. Using this mapping leads to the following forms for single and double excitation operators:

\begin{equation}
\hat \tau_{i}^{a} = t_i^{a}(a^\dagger_i a_a - a^\dagger_a a_i) = \frac{it_{i}^{a}}{2} \bigotimes_{k=i+1}^{a-1} Z_k(Y_iX_a - X_iY_a),
\end{equation}
\begin{align}
\begin{split}
\hat \tau_{ij}^{ab} &= t_{ij}^{ab}(a^\dagger_a a^\dagger_b a_i a_j - a^\dagger_j a^\dagger_i a_a a_b)  \\
 &=\frac{it_{ij}^{ab}}{8}  \bigotimes_{k=i+1}^{j-1} Z_k  \bigotimes_{l=a+1}^{b-1} Z_l (X_i X_j Y_a X_b + Y_i X_j Y_a Y_b \\
 &+X_i Y_j Y_a Y_b + X_i X_j X_a Y_b  - Y_i X_j X_a X_b \\
 &- X_i Y_j X_a X_b - Y_i Y_j Y_a X_b - Y_i Y_j X_a Y_b).
\end{split}
\end{align}
The exponential of each of these strings of Pauli matrices can be computed using a set of gates known as a Pauli gadget,\cite{Cowtan2020} a controlled version of which is shown in \Cref{fig:controlled_pauli}.

 As there is no immediate expectation-value form similar to \Cref{eq:exp_resid} for $r_\mathbf{i}^\mathrm{u}$ and $s_\mathbf{i}$ computing such quantities requires auxiliary qubits. Two alternative circuit constructions to obtain $r_\mathbf{i}^\mathrm{u}$ and $s_\mathbf{i}$ are given in \Cref{fig:circs}. 

The first uses fewer ancilla qubits, but it also requires a controlled version of the unitary $\hat U$, which decomposes into a larger number of two-qubit gates than its uncontrolled couterpart, regardless of the nature of $\hat U$. As a significant fraction of NISQ device error is due to two-qubit gates, the second circuit may be preferable on hardware. However, { it requires an ancilla register of the same size as the system register, which causes a $2^{n_\mathrm{qubits}}$-fold increase of the size of the qubit Hilbert space, leading to worse performance on quantum simulators}. Work presented heretherefore employs the circuit in \Cref{fig:circs} (a). 

In the particular case of the Pauli gadget-based unitary coupled cluster Ansatz, controlled-$\hat U$ only requires one additional controlled rotation gate per Pauli gate string, making it relatively easy to implement. { Compared to a given UCC circuit employed in conventional VQE, this approach requires one additional qubit and $n_\mathrm{electrons} + 2$ additional gate layers. The method also requires a constant number of additional two-qubit gates for each parameter in the Ansatz. The asymptotic scaling of circuit depth and two-qubit gate count is therefore equivalent to a conventional VQE algorithm employing the same UCC Ansatz.

Labelling the overall state in the qubit register of the circuit in \Cref{fig:circs} (a) as $\ket{\Phi}$, the expectation value of $Z$ on the ancilla is given by
\begin{widetext}
\begin{equation}
        \braket{\Phi|\hat Z_\mathrm{anc}|\Phi} =\frac{1}{4} (\braket{\phi_i| \hat U\phi_0} + \braket{\phi_0 \hat U^\dagger|\phi_i} + \braket{\phi_i| \hat U \phi_0} + \braket{\phi_0 \hat U^\dagger|\phi_i}
        = \frac{1}{4} (4 \times \Re{\braket{\phi_i|\hat U\phi_0}}) = \Re{\braket{\phi_i|\hat U\phi_0}},
\end{equation}
\end{widetext}
so $\braket{\Phi|\hat Z_\mathrm{anc}|\Phi} = \Re{(s_\mathrm{i})}$. The $Y$-expectation value similarly gives the imaginary part of the overlap. Additionally, applying the Hamiltonian to the system register and measuring the expectation value gives the corresponding parts of the residual,
\begin{equation}
\braket{\Phi|\hat Z_\mathrm{aux}\hat H|\Phi} = \Re{\braket{\phi_i|\hat H|\hat U\phi_0}} = \Re{(r^u_\mathbf{i})}.
\label{eq:rri}
\end{equation}

{In order to do this, the second-quantised Hamiltonian in \Cref{eq:ferm_hamil} must be mapped onto a corresponding qubit Hamiltonian, which can be done using the same techniques discussed for excitation operators. This leads to an operator of the form
\begin{equation}
    \hat H = \sum_k h_k \hat P_k,
    \label{eq:qubit_ham}
\end{equation}
where $\hat P_k$ are strings of Pauli operators. The expectation value in \Cref{eq:rri} can then be rewritten as
\begin{equation}
 \braket{\Phi|\hat Z_\mathrm{aux}\hat H|\Phi} = \sum_k h_k\braket{\Phi|\hat Z_\mathrm{aux}\hat P_k|\Phi}   
\end{equation}
and each term measured individually. The cost of measuring each residual therefore scales with the number of terms in the Hamiltonian, like the total energy calculation.} Residuals computed in this way can be employed in a projector method based on \Cref{eq:param_resid}, leading to an unlinked PQE algorithm. Given noiseless, exact values for $r_\mathbf{i}^\mathrm{u}$ and setting the shift to track the instantaneous projected energy, such an algorithm would converge directly to the ground state, without need for further Monte Carlo estimation. { However, expectation values obtained from a realistic quantum device are subject to at least finite-sampling noise. Therefore, it is perhaps wiser to consider the measured values of $r_\mathbf{i}^\mathrm{u}$ as stochastic estimates of the underlying quantities and therefore use them in a QMC-like algorithm. The resulting MC-PQE method is outlined below}:
\begin{enumerate}
    \item On a classical machine, represent the wavefunction $\hat U (\boldsymbol{\theta})\ket{\phi_0}$ as a distribution of walkers with populations $N_0$ corresponding to $\ket{\phi_0}$ and $N_i = \theta_i N_0$ corresponding to each parameter.
    \item Encode the wavefunction $\hat U (\boldsymbol{\theta})\ket{\phi_0}$ on a quantum device and measure the residuals $r^u_\mathrm{i}$. { This is done here using the circuit in \Cref{fig:circs} (a). Measuring each residual is equivalent to an energy measurement. As in VQE, this requires individual measurement of each (non-commuting) term in \Cref{eq:qubit_ham}, scaling as $n_\mathrm{qubit}^4$. The number of residuals is equal to the number of parameters in the employed Ansatz.}
    \item Update each $N_\mathbf{i} = N_\mathbf{i} - N_0 \times \delta\beta \times r_\mathbf{i}$.
    \item Store the projected energy $E_\mathrm{proj} = r_0 + S * s_0$.
    \item Once a threshold population $N_\mathrm{tot} = \sum_{\mathbf{i}=0}^{N_\mathrm{param} +1} N_\mathbf{i}$ has been reached, allow the shift to vary to maintain this population constant. In { the original FCIQMC algorithm }this is done by\cite{Booth2009}
    \begin{equation}
        S(\beta) = S(\beta - A\delta\beta) - \frac{\zeta}{A\delta\beta}\mathrm{ln}\frac{N_\mathrm{tot}(\beta)}{N_\mathrm{tot}(\beta - A\delta\beta)},
        \label{eq:shift}
    \end{equation}
    { although more recent update procedures have been developed that lead to finer population control and more rapid convergence.\cite{Yang2020}}
    \item Repeat steps 2-5 until sufficient samples of $E_0$ and $S$ have been accumulated to obtain accurate Monte Carlo estimates.
\end{enumerate}
On the one hand, as discussed in Sec. \ref{sec:res}, this approach has a series of benefits relative to a ``deterministic" PQE approach, as it is resilient to noise in the quantum measurement, allowing the use of low shot numbers to obtain accurate results. Additionally, the shift provides a second independent estimator for the energy which can be used to more clearly identify when convergence has been reached. On the other hand, it is more efficient than a fully classical Quantum Monte Carlo algorithm as it both removes the need for cluster sampling and merges the death and spawning steps into a single process.

One of the main benefits of QMC methods over conventional algorithms is that they take advantage of sparsity in the Hamiltonian to store a compressed version of the wavefunction, thereby decreasing classical memory overheads. In the following section we explore these QMC-inspired ideas to further decrease the memory or quantum resource costs of the algorithm detailed above.

\subsection{Stochastic approximations}

\subsubsection{Sampling the Hamiltonian}
\label{sec:hamil}
 When computing the residuals $r_\mathbf{i}^\mathrm{u}$ one must in principle measure each of the terms of the qubit Hamiltonian separately, as they require different post-rotations to be applied to the circuit before measurement. However,  they can be further grouped\cite{McClean2016, Kandala2017} as 
\begin{equation}
    \hat H = \sum_k \big( \sum_{j \in G_k}  h_{kj} \hat P_{kj}\big), 
\end{equation}
where $[\hat P_{ki}, \hat P_{lj}] \propto \delta_{kl}$. Terms in the same group $G_k$ commute, so they may be measured simultaneously using the same post-rotations. { Nevertheless, in the general case, this does not reduce the operator to a single group, so a significant number of terms still need to be measured independently. See the caption of \cref{tab:hydrides} for some examples.} Additionally, optimal grouping of Pauli terms is in general a hard problem.\cite{Gokhale2020} For the two-body quantum chemistry Hamiltonian there is a simple heuristic partitioning, based on the qubitwise commutativity of Pauli strings. This leads to one large group comprising the identity string and all strings containing only $Z$ gates and many small groups of strings containing some $X$ and $Y$ gates. { Significant work has gone into further optimising the number of measurements required to obtain Hamiltonian expectation values, including modifying the partitioning between groups\cite{Choi2023, Wu2023,Yen2023} or the Pauli strings themselves\cite{Choi2022} to lower variance, or using (randomised) measurement techniques based on the classical shadows approach.\cite{Huang2020,Hadfield2021, Huang2021, Choi2023b}}

Regardless of the means by which groups are obtained, one can employ an approach similar to that used in classical QMC algorithms and only select a subset of the groups to measure at each time-step. This decreases the number of circuit measurements needed per time-step, but introduces further noise in the estimate of $r_\mathbf{i}^\mathrm{u}$. However, provided no systematic bias is caused, this noise will be averaged out over the course of the imaginary-time propagation.

There exist various possible Hamiltonian term selection schemes, the simplest of which would be to sample uniformly from the set of Pauli terms or groups. Many of these terms will have negligible weights however, so in order to better capture the action of the Hamiltonian in a small number of samples, some form of importance sampling is recommended. One straightforward approach is to select each Pauli group with probability proportional the group weight $W_k = \sum_{j \in G_k} |h_{kj}|$,

\begin{equation}
    p(k) = \frac{W_k}{\sum_k W_k}
\end{equation}

For most quantum chemistry Hamiltonians investigated in this paper, we find that, when using qubitwise commutativity to partition the terms,  the weight of the group containing the empty Pauli string and all $Z$-only strings, which we will label $G_0$, strongly dominates the resulting probability distribution. Therefore, under this selection scheme, it is likely that for small numbers of samples it will be \textit{over-} rather than \textit{under}sampled. We also note that this term is qualitatively different from the other groups, as it describes the diagonal action of the Hamiltonian, while the others all correspond to off-diagonal contributions. Therefore, to obtain a more balanced description, we propose a scheme in which the diagonal group $G_0$ is always evaluated, while the other groups are selected with probability
\begin{equation}
    p'(k) = \frac{W_k}{\sum_{k\neq 0} W_k}
\end{equation}
\subsubsection{Sampling the wavefunction}

UCC is the natural candidate Ansatz for this type of quantum-enhanced QMC algorithm. One of the main challenges with using this Ansatz on NISQ devices is that it leads to deep circuits, with execution times that easily exceed the coherence times for current machines. We have previously shown that by using insight from a fully classical UCCMC calculation,\cite{Filip2020} one can significantly shorten circuits while preserving accuracy by removing low-weight excitors from the Ansatz.\cite{Filip2022}

This circuit simplification can be done on-the-fly during the propagation, by stochastically rounding the parameter values before encoding the wavefunction,
\begin{equation}
    \widetilde \theta_\mathbf{i} = \begin{cases}
    \theta_\mathbf{i},\ \mathrm{if}\ \frac{\sum_\mathbf{j}^\mathbf{i}\theta_\mathbf{j}}{\sum_\mathbf{j}^{\mathbf{i}_\mathrm{max}}\theta_\mathbf{j}} \leq p\\
    0,\ \mathrm{otherwise},
    \end{cases}
\end{equation}
where $p$ is a uniform random sample from the interval $[0,1)$ and the parameters $\theta_\mathbf{i}$ are ordered by decreasing amplitude. Once again, this is expected to increase fluctuations in the estimators at each time-step. The issue of potential systematic biases is explored further in Sec. \ref{sec:res}.

\subsubsection{Spawning}\label{sec:spawn}
In the methodology detailed above, all parameters are updated at every time-step. However, one can take the approach commonly employed in QMC algorithms and only update a stochastically selected subset of the populations. In particular, for spawning in conventional QMC algorithms one can use a variety of excitation generation algorithms.\cite{Holmes2016,Neufeld2019,Gunther2020,Weser2023} 

In principle, one could spawn onto connected determinants uniformly. However the spawning probability is proportional to $\frac{H_\mathbf{ij}}{p_\mathrm{gen}(\mathbf{i|j}) }$ (see \Cref{eq:pspawn}), so for appropriate importance sampling of this process it is desirable that ${p_\mathrm{gen}(\mathbf{i|j}) } \propto {H_\mathbf{ij}}$. Different algorithms have been designed to approximately achieve this, using heat-bath sampling\cite{Holmes2016,Weser2023} or mathematical bounds on the value of ${H_\mathbf{ij}}$.\cite{Neufeld2019}

With the residuals computed according to the quantum algorithm in Sec. \ref{sec:algorithm}, death and spawning are no longer separate processes, so the parameters to be updated should be generated with 
{ 
\begin{equation}
{p_\mathrm{gen}(\mathbf i) } \propto |\braket{\phi_\mathbf{i}|\hat H - S |\Psi}|.
\end{equation}

{ If we write out the fermionic Hamiltonian as
${\hat H = \sum_k h^\mathrm{F}_k \hat F_k}$, where $\hat F_k$ is some fermionic excitation operator and (non-uniquely) label the excitation between determinants $\phi_\mathbf{i}$ and $\phi_\mathbf{j}$ as $\hat F_\mathbf{ij}$, then 
\begin{equation}
    {p_\mathrm{gen}(\mathbf i) } \propto |\sum_\mathbf{j} c_\mathbf{j}h^\mathrm{F}_\mathbf{ij}|
    \label{eq:prob1}
\end{equation}
Sampling the distribution $p_\mathrm{gen}$ is non-trivial even with access to a quantum computer and a qubit representation of the wavefunction and Hamiltonian.} In particular, the circuits described in \Cref{fig:circs}, which can be used to estimate the overlap itself, cannot be used for this purpose. 

However, one can efficiently generate determinants with probability  
\begin{equation}
    \tilde{p}_\mathrm{gen}(\mathbf i) \propto \sum_\mathbf{j}|c_\mathbf{j}^2h^\mathrm{F}_\mathbf{ij}|
    \label{eq:prob2}
\end{equation}
by preparing the state $\ket{\Psi}$, applying one of the operators $\hat P_k$ corresponding to each $\hat F_k$ to it with probability $\frac{|h^\mathrm{F}_k|}{\sum_k|h^\mathrm{F}_k|}$ and then measuring the qubit register. While this is not a perfect importance sampling of the Hamiltonian coupling, it will better match its minima and maxima and should therefore represent an improvement over uniform sampling. An example for the H$_4$ molecule is shown in \Cref{fig:probs}.} These circuits do not require ancilla or additional controlled operations, so they are less costly to implement than those needed to compute the residuals. The number of shots needed is equal to the number of determinants one wants to generate, so they would add little computational overhead.

\begin{figure}
    \centering
    \includegraphics[width = 0.45\textwidth,, trim = 1.5cm 0.5cm 3cm 1.5cm, clip]{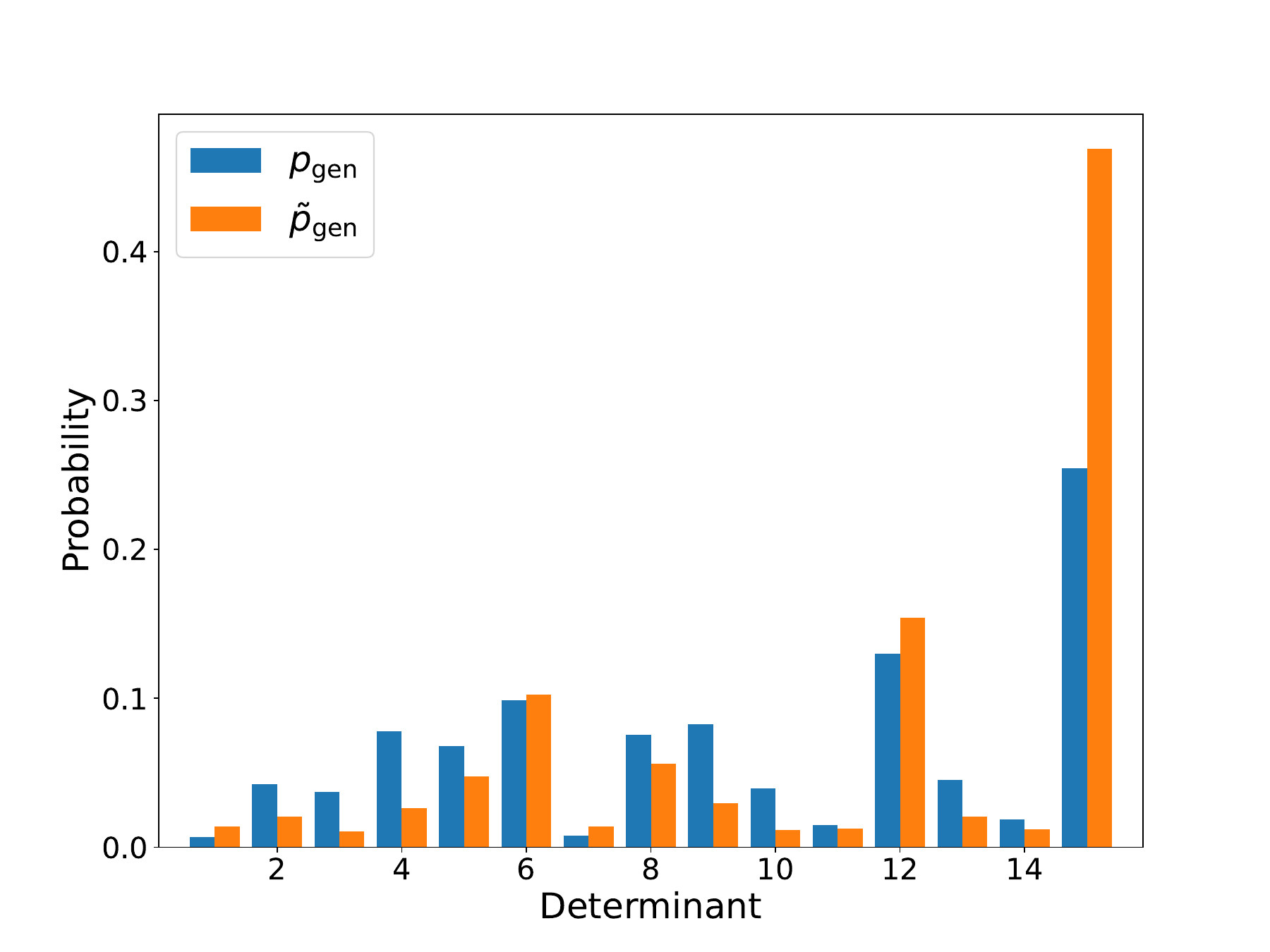}
    \caption{\small \justifying  Excitation generation probability according to \Cref{eq:prob1} and \Cref{eq:prob2} for the H$_4$ molecule, using an approximate ground state wavefunction. $\tilde{p}_\mathrm{gen}$ is a better approximant for ${p}_\mathrm{gen}$ than the uniform distribution. As a metric, the Kullback–Leibler divergence\cite{Kullback1951} of $\tilde{p}_\mathrm{gen}$ with respect to ${p}_\mathrm{gen}$ is 0.118 , while that of the uniform distribution is 0.541.}
    \label{fig:probs}
\end{figure}
While stochastic parameter updates may be valuable for larger systems, for the problems considered in this paper it is easy to store and cycle through all the relevant parameters and therefore probabilistic spawning is not employed.

\subsection{Folded spectrum considerations}
All the methods described above can equally be applied to the folded spectrum Hamiltonian, by replacing the residual with that in \Cref{eq:res_fs}. As the squared Hamiltonian operator contains many-body terms of higher order than the original Hamiltonian, this will lead to more terms in the Pauli string representation of the corresponding qubit operator as well. However, some of these will commute with the terms in the original operator, so the number of different groups will see a smaller increase. { Nevertheless, for non-trivial systems, this leads to a potentially insurmountable increase in required computation, even with efficient Hamiltonian term grouping methods. The possibility of sampling the Hamiltonian is therefore particularly promising in this case as a means to mitigate the additional costs.}

\begin{figure}
    \centering
    \includegraphics[width=0.5 \textwidth, trim = 1cm 0.7cm 3cm 1.5cm, clip]{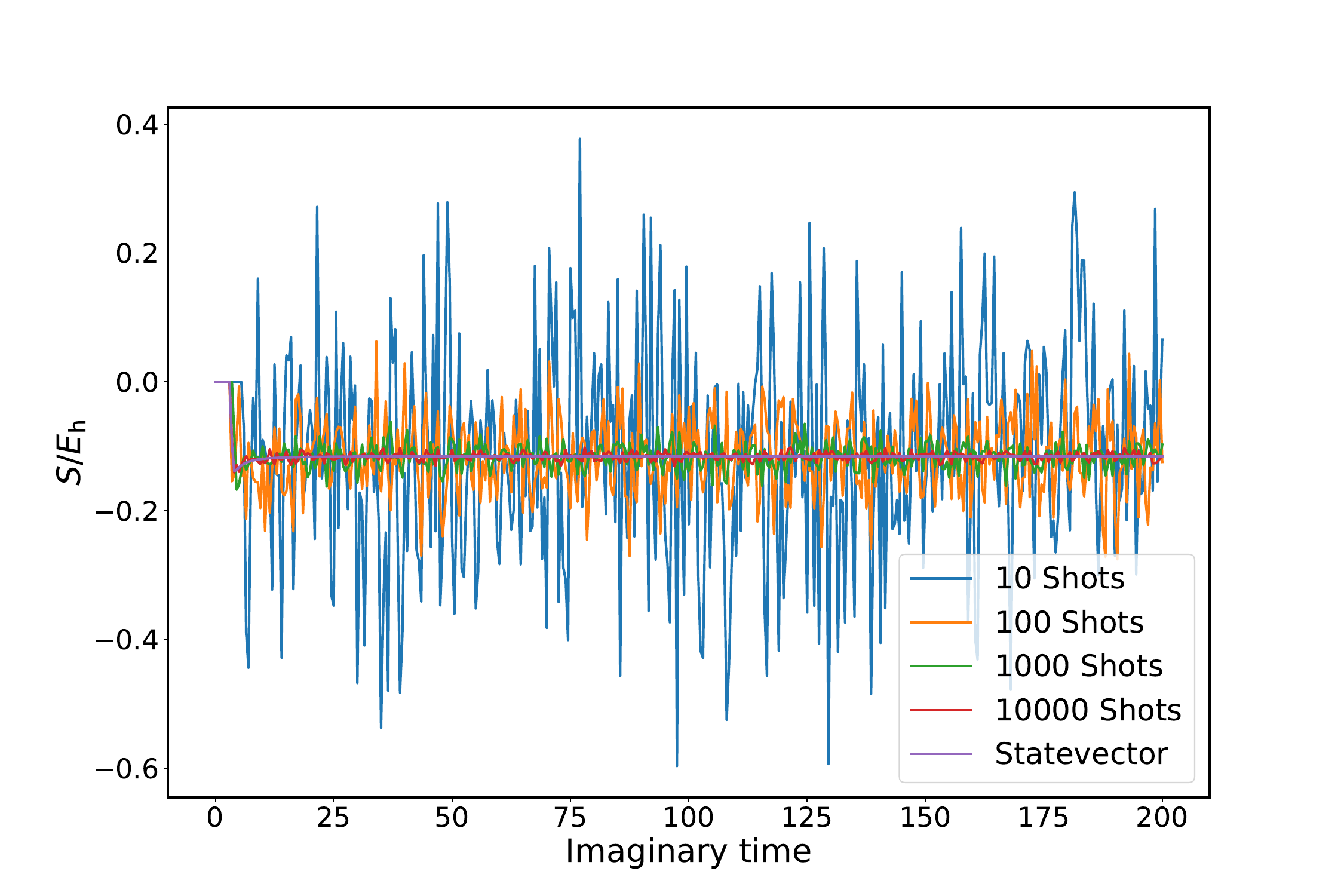}\\
    \includegraphics[width=0.5 \textwidth, trim = 1cm 0.7cm 3cm 2cm, clip]{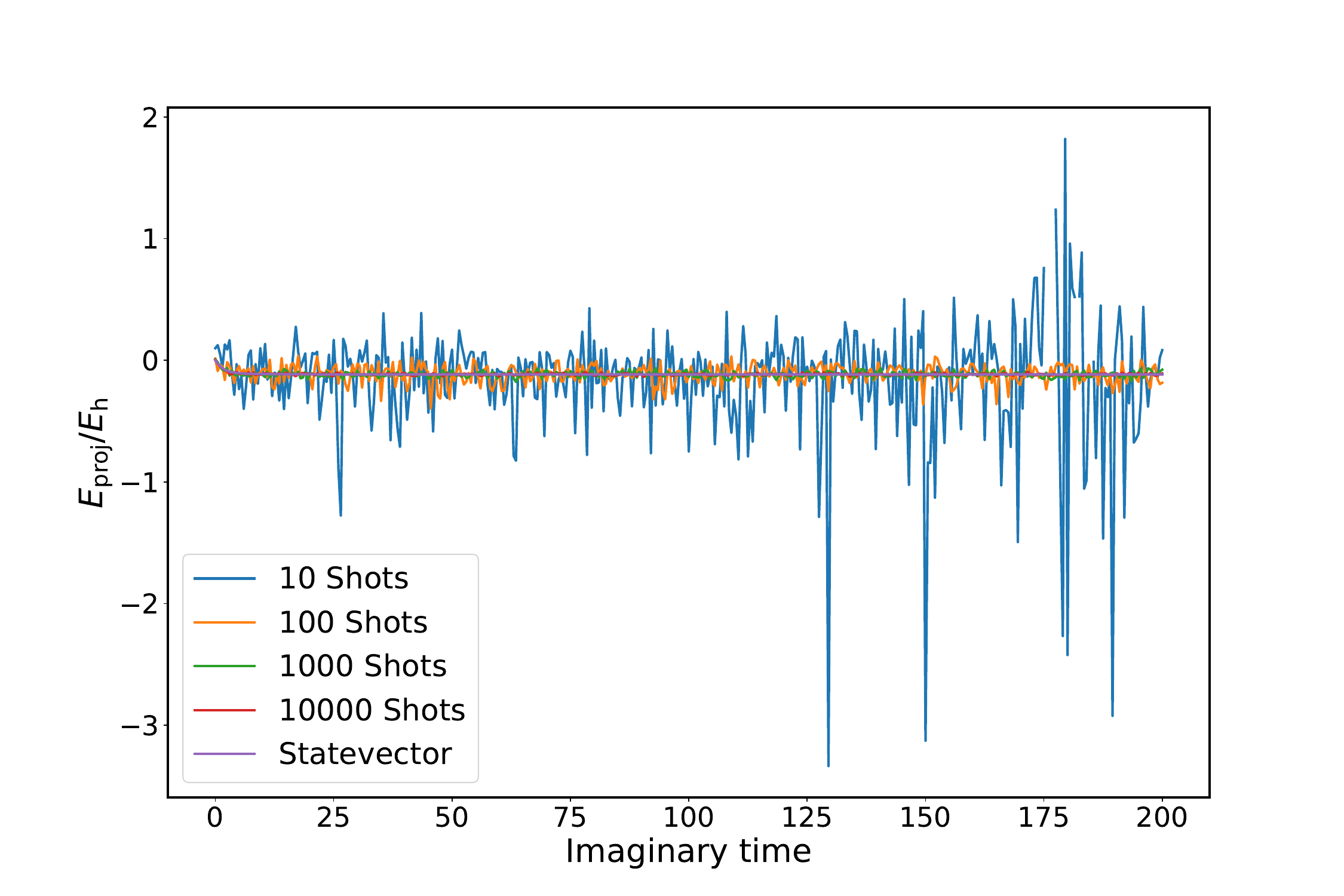}
    \caption{\small \justifying Shift (top) and projected energy (bottom) as a function of imaginary time, obtained from $10^1 - 10^4$ samples of the residual circuits for each Pauli term in the Hamiltonian,  for H$_3^+$ at $r = 2.0$ \r{A}. Noise decreases as the number of shots increases.}
    \label{fig:shot_noise}
\end{figure}

As molecules are pulled apart electronic states approach one another and often become degenerate in the fully dissociated limit. Therefore, if one is searching for two different states which are close in energy (which will remain so after the folding procedure), noise in the wavefunction representation becomes potentially problematic as it may be sufficient to induce a switch from one state to another. { In the case of ground state calculations, it has been observed in conventional CCMC that, particularly in more strongly correlated regimes,   large fluctuations will sometimes cause a calculation to collapse onto a different, often unphysical, state.\cite{Scott2019,Scott2020} This is made possible by the nonlinear structure of the coupled cluster equations, which allow for multiple solutions, not all of which have a configuration interaction correspondent.\cite{Piecuch2000,Csirik2023}} Therefore, the range of stochastic techniques that can be used in this scenario may be more limited than in ground state calculations.

\section{Results}\label{sec:res}
All results in this paper are obtained from state-vector or shot-based quantum circuit simulation using the t$\ket{\mathrm{ket}}$ platform.\cite{Sivarajah2021} Necessary atomic orbital Hamiltonian integrals were obtained from PySCF\cite{Sun2020} and FCI calculations were performed using HANDE-QMC.\cite{Spencer2019} All MC-PQE calculations used the UCCSD Ansatz. The main systems considered are the H$_2$, H$_3^+$, H$_3$ and H$_4$ linear molecules in the STO-3G basis set,\cite{Hehre1969} which serve as a useful proof of concept for the methods described here. These molecules are small enough that the corresponding Fock space vectors can be easily manipulated on a classical computer, while covering a range of chemical behaviour. Shot-based simulations are obtained by taking $N_\mathrm{shots}$ samples from the probability distribution corresponding to the wavefunction encoded in a given quantum circuit. 
\begin{table}
    \centering
    \caption{\small \justifying Mean and standard deviation of the shift and projected energy as the number of shots is increased, for H$_3^+$ at $r = 2.0$ \r{A}. {All energies are given in Hartree.}}
    \label{tab:variance}
    \begin{tabular}{c|c|c|c|c|c}
       $N_\mathrm{shots}$  &  $\braket{S}$ & $\sigma (S)$ & $\braket{E}$ & $\sigma(E)$&$E_\mathrm{FCI}$\\
       \hline
         10 & -0.111 & 0.009 & - & - &\multirow{4}{*}{-0.11586}\\
         100 & -0.116& 0.003& -0.114& 0.004& \\
         1000 & -0.1167& 0.0009 & -0.117& 0.001& \\
         10000 & -0.1163 & 0.0003& -0.1155 & 0.0004& \\
         \hline
    \end{tabular}

\end{table}
In order to reduce computation time required for larger systems, shot-based simulation is at times replaced by state-vector simulation with added Gaussian noise of mean $\mu = 0$ and standard deviation $\sigma$. The average behaviour of such simulations should be very similar to true shot-based results for large enough $N_\mathrm{shots}$, but it will fail to capture some of the catastrophic failures of the very low shot count regime.

Average values of the shift $S$ and $E_\mathrm{proj}$ are obtained by a reblocking analysis\cite{Flyvbjerg1989} of data obtained in the steady-state regime of calculations, using pyblock.\cite{pyblock}

\subsection{Ground State Calculations}

We begin our investigation into the effects of sampling noise on the MC-PQE algorithm by carrying out calculations using the circuit in \Cref{fig:circs} (a), sampled with { $10^1\ \mathrm{ to}\ 10^4$} shots for each Pauli term in the Hamiltonian. 

Representative { examples of MC-PQE energies over the course of calculations with differing numbers of shots} for H$_3^+$ at a bond length of $r = 2.0$ \r{A} are shown in \Cref{fig:shot_noise}. { The instantaneous energy estimators display significant fluctuations, although, unsurprisingly, the noise decreases as the number of shots is increased. There is a qualitative change in the behaviour of the projected energy using $N_\mathrm{shots} = 10$, with some samples deviating significantly from the expected mean. This is effectively due to the underestimation of the denominator in the expression of $E_\mathrm{proj}$ (see \Cref{eq:e_proj}), which leads to divergences in this quantity. There is therefore a lower bound on the number of shots needed to obtain a stable calculation, which is dependent on the overlap $\braket{\phi_0|\Psi(\beta)}$. The final estimators of the energy, obtained by averaging over the instantaneous energies once the calculation has reached its steady-state regime are given in \Cref{tab:variance} and display the expected lower variances as the number of shots increases.}  

\begin{figure}[h]
    \centering
    \includegraphics[width=0.45 \textwidth, trim = 0cm 0.5cm 3cm 1.5cm, clip]{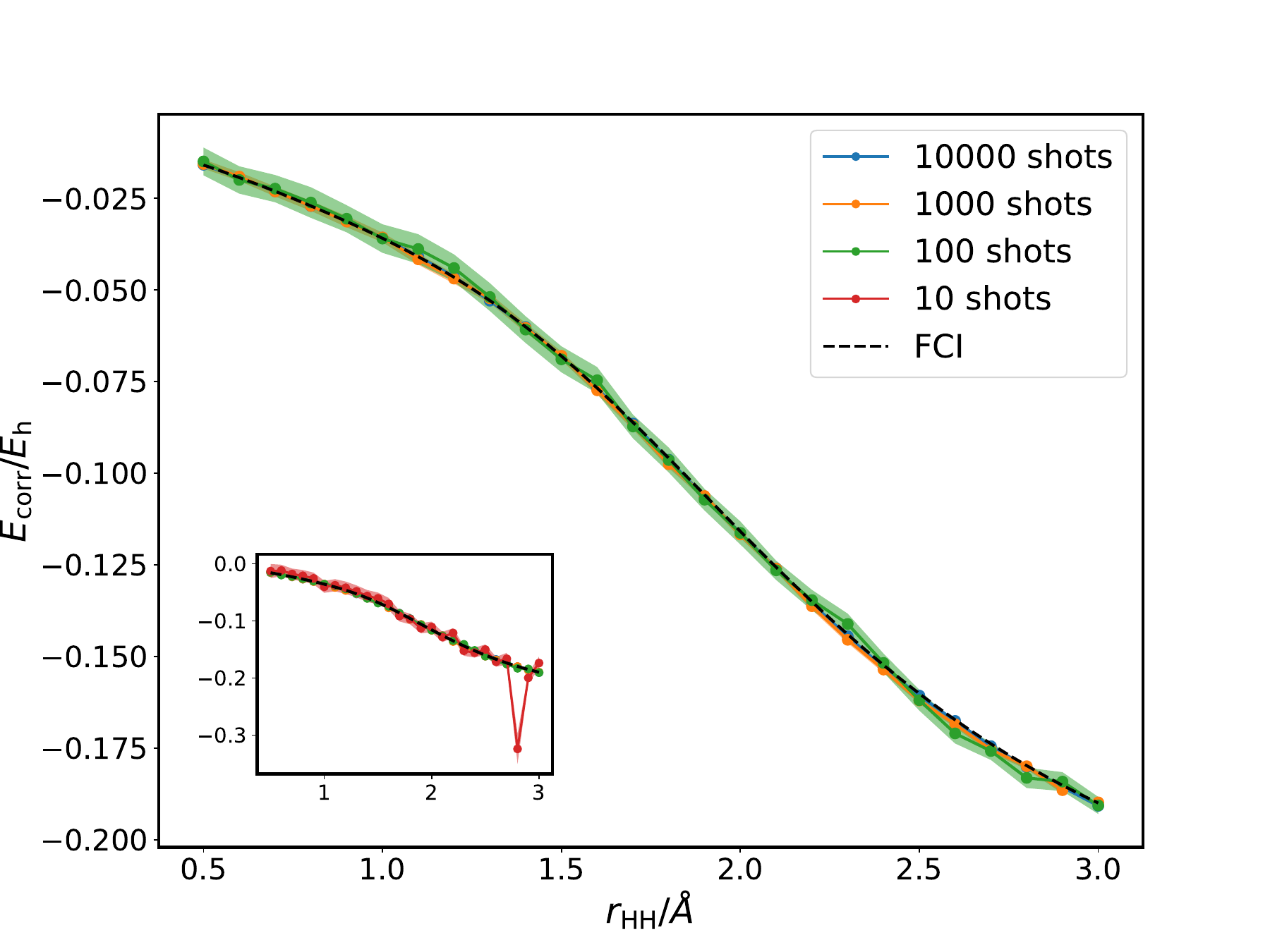}
    \includegraphics[width=0.45 \textwidth, trim = 0cm 0.5cm 3cm 2cm, clip]{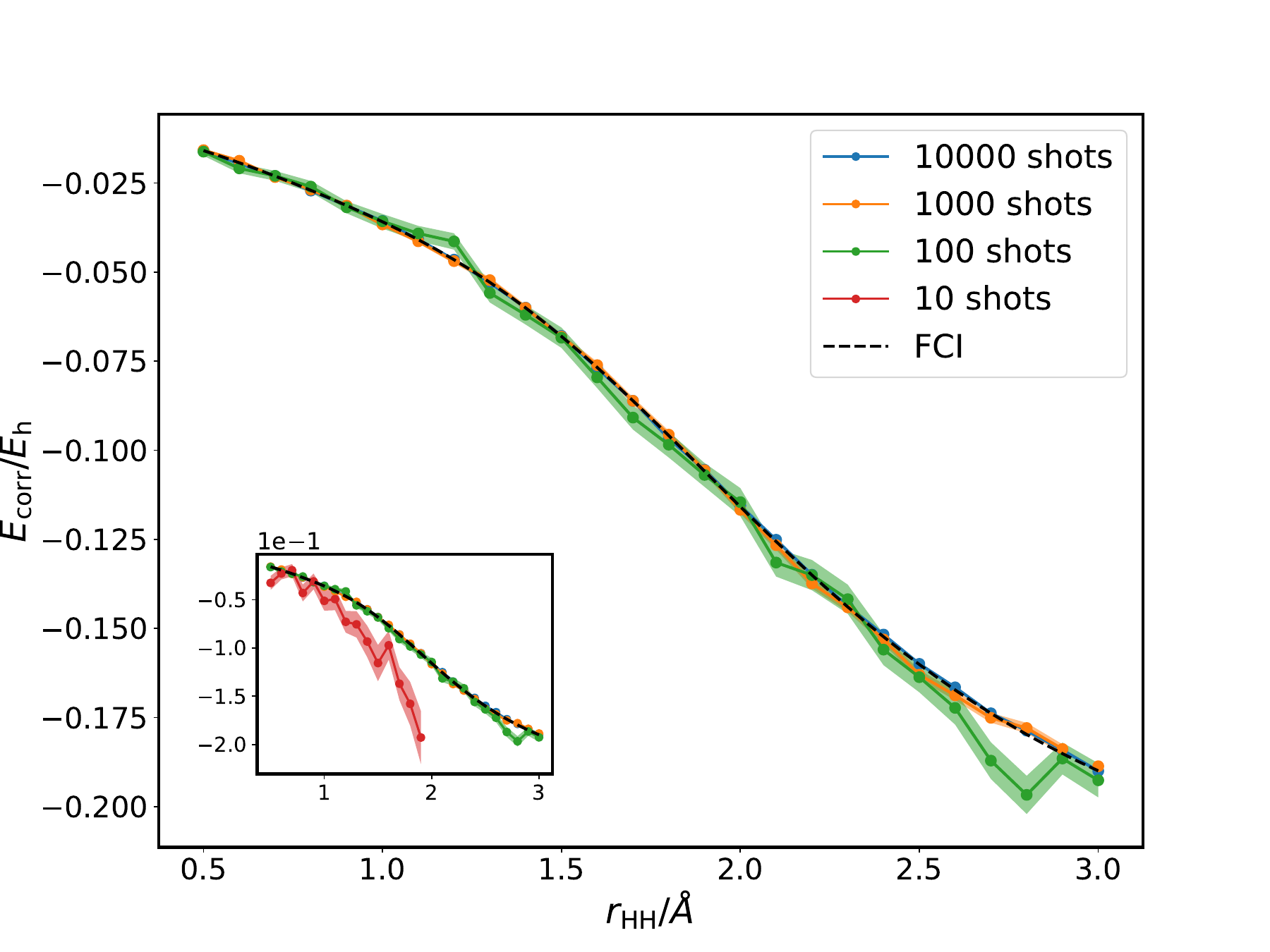}
    \caption{\small \justifying Shift and $E_\mathrm{proj}$ estimates of the correlation energy of linear H$_3^+$ over the range $r_\mathrm{HH} = 0.5 - 3.0 a_0$. Deviation from the true correlation energy increases as the number of shots is decreased, with deviations more pronounced in the projected energy and at larger bond lengths. This correlates with an increase in strong correlation and lowering of the overlap of the ground state wavefunction and the Hartree--Fock reference, $\braket{\phi_0|\Psi(\beta)}$.} 
    \label{fig:binding}
\end{figure}

\begin{table}[h]
    \centering
    \caption{\small \justifying Mean and standard deviation of the shift and projected energy obtained from the deterministic { statevector} propagation of stochastically rounded wavefunctions for systems of increasing size. { All systems have bond lengths of $r = 1.5$ \r{A} and all energies are given in Hartree.} }
    \label{tab:stoch_wfn}
    \begin{tabular}{c|c|c|c|c|c}
      System &  $\braket{E}$ & $\sigma(E)$ & $\braket{S}$ & $\sigma(S)$&$E_\mathrm{FCI}$\\
      \hline
         H$_3^+$& -0.068 & 0.001 & -0.0688&0.0005&-0.06803 \\
         H$_3$ & -0.1107& 0.0009& -0.113& 0.001&-0.11011\\
         H$_4$ & -0.176& 0.002& -0.180& 0.003& -0.16701\\
         \hline
    \end{tabular}

\end{table}

\begin{figure}[h!]
    \centering
    \includegraphics[width=0.45 \textwidth, trim = 1cm 0.5cm 3.1cm 2cm, clip]{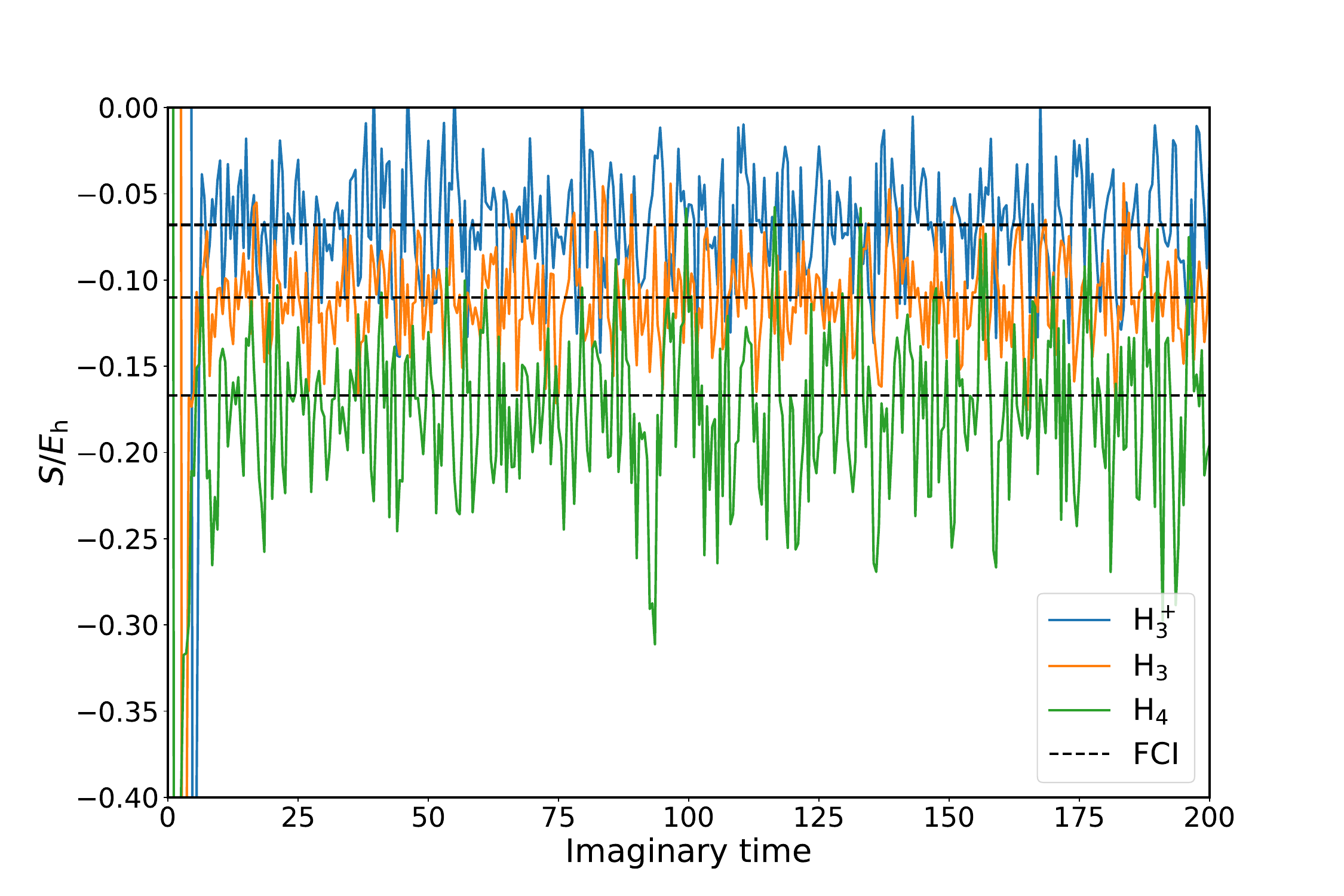}
    \includegraphics[width=0.45 \textwidth, trim = 1cm 0.5cm 3.1cm 2cm, clip]{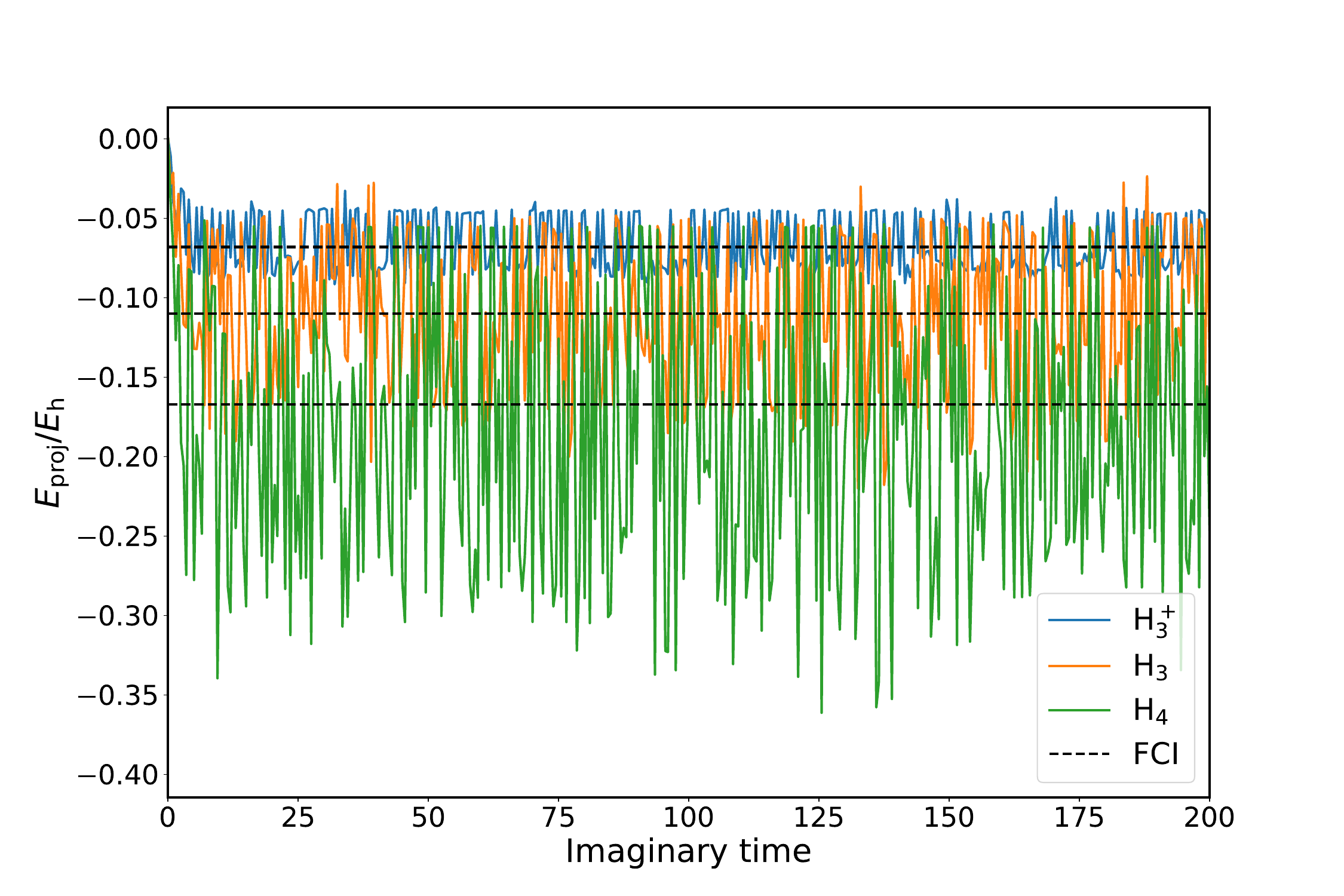}
    \caption{\small \justifying Shift (top) and projected energy (bottom) as a function of imaginary time, obtained from a stochastically represented wavefunction for H$_3^+$ (blue), H$_3$ (orange) and H$_4$ (green) at $r = 1.5$ \r{A}. The noise in the shift is roughly system-independent, whereas the variance of $E_\mathrm{proj}$ increases with its absolute value.}  
    \label{fig:stoch_wfn}
\end{figure}

\begin{table*}[]
    \centering
    \caption{\small \justifying  Comparison of VQE and MC-PQE results for H$_3^+$ and H$_4$. $n_\mathrm{2q}$ gives the number of two-qubit gates in each state preparation circuit. Each method was run for $n_\mathrm{iter}$ optimisation cycles, using $n_\mathrm{shots}$ measurements per Pauli string measured at each step. $n_\mathrm{total}$ gives the total number of measurements required for the calculation. For VQE with SPSA optimisation, this is $2n_\mathrm{iter}n_\mathrm{hamil}n_\mathrm{shots}$, while for MC-PQE it is $(n_\mathrm{param} + 1)n_\mathrm{iter}n_\mathrm{hamil}n_\mathrm{shots}$. $\Delta E$ is the average absolute error over the binding curve, given in milliHartree. For a fair comparison, a variational estimate of the energy is also computed for MC-PQE.}
    \label{tab:comp}
    \begin{tabular}{c|c|c|c|c|c|c|c|c|c|c|c|c}
         \multirow{2}{*}{Method} & \multicolumn{6}{c|}{H$_3^+$} & \multicolumn{6}{c}{H$_4$}\\
         \cline{2-13}
         & $n_\mathrm{qubits}$ & $n_\mathrm{2q}$ & $n_\mathrm{shots}$ & $n_\mathrm{iter}$ & $n_\mathrm{total}$ & $\Delta E$ & $n_\mathrm{qubits}$ & $n_\mathrm{2q}$ & $n_\mathrm{shots}$ & $n_\mathrm{iter}$ & $n_\mathrm{total}$ &$\Delta E$ \\
         \hline
         VQE &  6 & 128 & 10000 & 1000 &  5.13e8  & 30 & 8 & 736 & 10000& 1000 & 2.07e9 & 43\\
         MC-PQE & 7 & 152& 100 & 250 & 3.13e6 & 0.7 & 9 & 832&100&250 &3.79e7& 1.5\\
         \hline
    \end{tabular}
    
\end{table*}

Average values for the estimators $S$ and $E_\mathrm{proj}$ along the binding curve of H$_3^+$ are given in \Cref{fig:binding}. At large bond-lengths, we observe that divergences in the instantaneous values of the estimators due to undersampling translate into divergences of the overall estimator, with the projected energy more sensitive to this than the shift. This behaviour is not unexpected. As the bond length increases, the system becomes increasingly strongly correlated and therefore the weight of the reference determinant $\phi_0$ decreases, requiring more sampling to estimate appropriately.

It is therefore the case that more strongly correlated systems will require more sampling than weakly correlated ones. This follows well-known trends in the difficulty of solving quantum chemical problems in a classical setting and we can offer a series of ways to alleviate the issue. Firstly, in the absence of noise, the overlap $\braket{\phi_0|\Psi(\beta)}$ decreases monotonically with $\beta$. Therefore, it will be possible to use fewer shots at short imaginary times and increase the shot count as the calculation progresses, thereby accelerating convergence to the steady state. For most Monte Carlo calculations however, the majority of the time is spent sampling this state, which cannot benefit from the early reduction in shots. 

\begin{table}
    \centering
    \caption{\small \justifying Mean and standard deviation of the shift and projected energy obtained from linearised imaginary time-propagation of the H$_3^+$ wavefunction at $r=2.0$ \r{A}, using stochastically sampled Hamiltonians with increasing numbers of terms selected per step. { All energies are given in Hartree.}}
    \label{tab:hamil}
    \begin{tabular}{c|c|c|c|c|c}
      $N_\mathrm{hamil}$ &  $\braket{E}$ & $\sigma(E)$ & $\braket{S}$ & $\sigma(S)$&$E_\mathrm{FCI}$\\
      \hline
         2&-0.125&0.007&-0.115&0.003&\multirow{5}{*}{-0.11586}\\
         3& -0.121& 0.004& -0.116& 0.001&\\
         4 & -0.121& 0.004& -0.117& 0.001&\\
         5 & -0.115& 0.004& -0.1137& 0.0007&\\
         6 & -0.112 & 0.003 & -0.1133& 0.0007&\\
         \hline
    \end{tabular}
    
\end{table}

Secondly, different numbers of shots can be allocated to different quantities of interest. In this case, one could run a calculation with a small number of shots for the residuals, but a large number of shots for the reference overlap estimation. As the system size increases, this will represent a smaller fraction of the computational effort needed, thereby allowing calculations that are effectively as shot-efficient as ones with the lowest number of shots.

Another standard computational chemistry approach would be to replace the Hartree--Fock reference wavefunction in the projected energy with a more complex, multi-determinental quantity $\ket{\Psi_\mathrm{T}}$ which is hoped to have larger overlap with the true ground state, giving a new projected energy of the form
\begin{equation}
    E'_\mathrm{proj} = \frac{\braket{\Psi_\mathrm{T}|\hat H| \Psi(\beta)}}{\braket{\Psi_\mathrm{T}|\Psi(\beta)}} = \frac{\sum_{\mathbf{i} \in \Psi_\mathrm{T}}c_{\mathrm{T}\mathbf{i}}\braket{\phi_\mathbf{i}|\hat H|\Psi(\beta)}}{\sum_{\mathbf{i} \in \Psi_\mathrm{T}}c_{\mathrm{T}\mathbf{i}}\braket{\phi_\mathbf{i}|\Psi(\beta)}}
\end{equation}

On a classical machine, this would require a linear increase in computation, however in the current scheme, all terms in both the numerator and denominator are already computed as part of the propagation. { We test this for $H_3^+$, which in a minimal basis has two electrons in three orbitals, using a trial wavefunction of the form 
\begin{equation}
    \ket{\Psi_T} = c_0\ket{01} + c_1\ket{23} + c_2\ket{45},
    \label{eq:trial_wfn}
\end{equation}
which includes the dominant configurations at dissociation, where $\ket{ij}$ is the Slater determinant with spin-orbitals $i$ and $j$ occupied. The orbitals are labeled in order of increasing energy, with even numbers corresponding to $m_s = 1/2$ and odd numbers to $m_s = -1/2$ orbitals. This is insufficient to ensure convergence of the extremely undersampled 10-shot calculation towards dissociation, but it does reduce the standard error in 100-shot calculations by a factor of approximately 1.5}

{ Finally we compare the performance of MC-PQE with conventional VQE, performed using the Simultaneous Perturbation Stochastic Approximation (SPSA) optimisation algorithm,\cite{Spall1992} which has been show to perform well for noisy variational optimisation.\cite{Bonet2023} In \Cref{tab:comp} we compare the average error in the ground state energy across the symmetric stretch of the $H_3^+$ and $H_4$ chains, obtained by VQE and MC-PQE. The performance of both methods could undoubtedly be improved by increasing the number of iterations and measurements, but as it stands for both of these systems MC-PQE can achieve  near chemical accuracy with at least two orders of magnitude fewer overall shots than conventional VQE.}

We move on to consider stochastic implementations of the projection algorithm itself. 
{We begin by considering the effects of stochastic sampling of the wavefunction and Hamiltonian independently, while doing the projection itself deterministically. Additional example MC-PQE trajectories are given in the appendix.} \Cref{fig:stoch_wfn} shows trajectories computed using deterministic projection on a stochastically rounded wavefunction for H$_3^+$, H$_3$ and H$_4$, with corresponding statistical properties given in \Cref{tab:stoch_wfn}. H$_2$ is not considered here as it only has one parameter in the wavefunction, so there is no rounding to be done. While the noise introduced into the shift by this procedure is roughly constant and system-independent, in itself a desirable property, the noise in $E_\mathrm{proj}$ scales linearly with the absolute magnitude of the correlation energy{, as can be seen from the bottom panel of \Cref{fig:stoch_wfn}}. This is unsurprising since if all amplitudes are rounded down, we expect the projected energy to be zero. Nevertheless, for all three systems we obtain stable estimators for the correlation energy. For the H$_3$ and H$_3^+$ systems, for which the UCCSD Ansatz is exact, the results are within error bars of the FCI value. For H$_4$, there is an error of ~10 milliHartree in the correlation energy estimators. This can be significantly reduced by computing the variational rather than projected energy estimator.\cite{Filip2020} 

\begin{figure}
    \centering
    \includegraphics[width=0.45 \textwidth, trim = 0 0 3cm 2.5cm, clip]{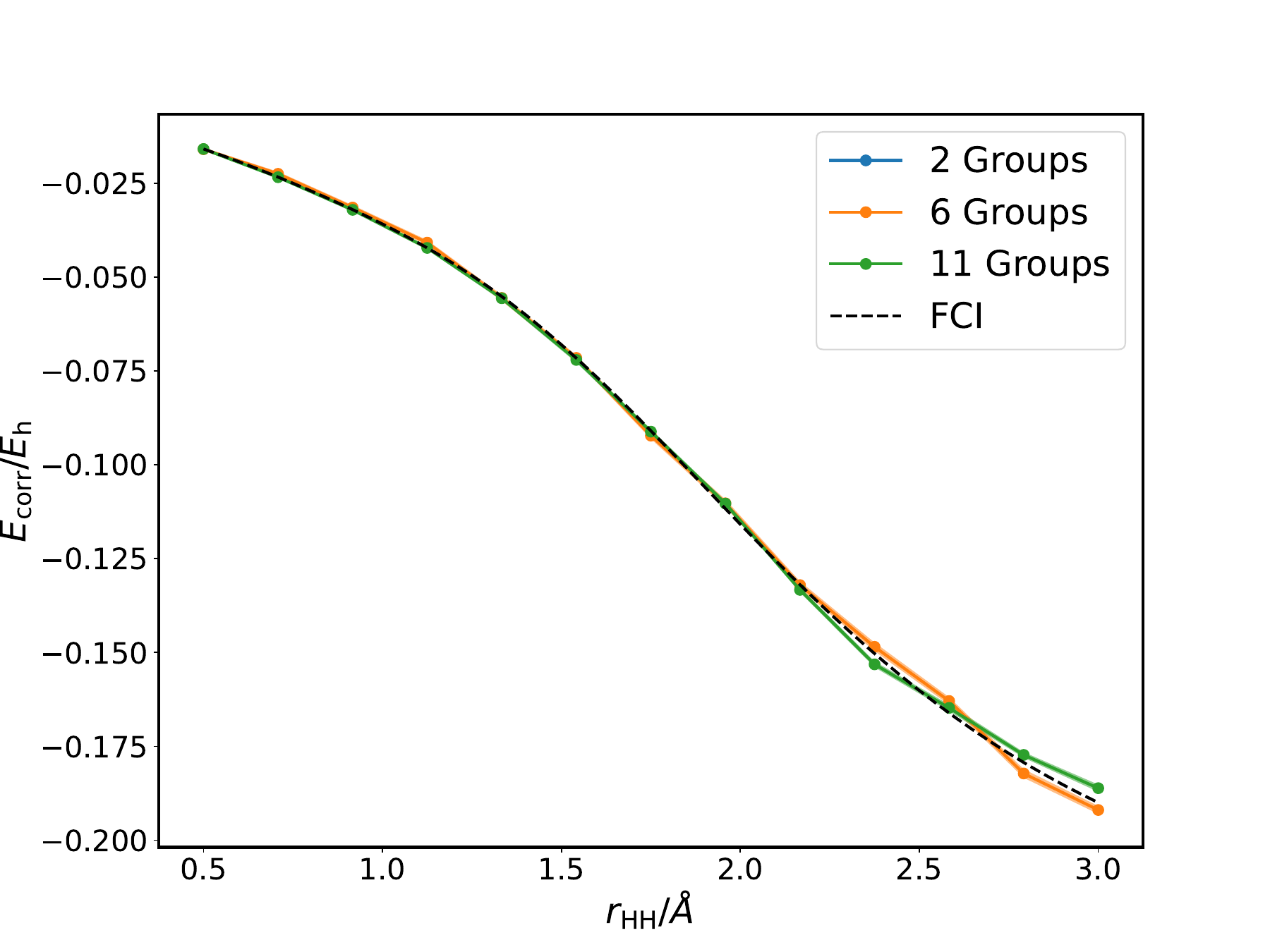}
    \includegraphics[width=0.45 \textwidth, trim = 0 0 3cm 2.5cm, clip]{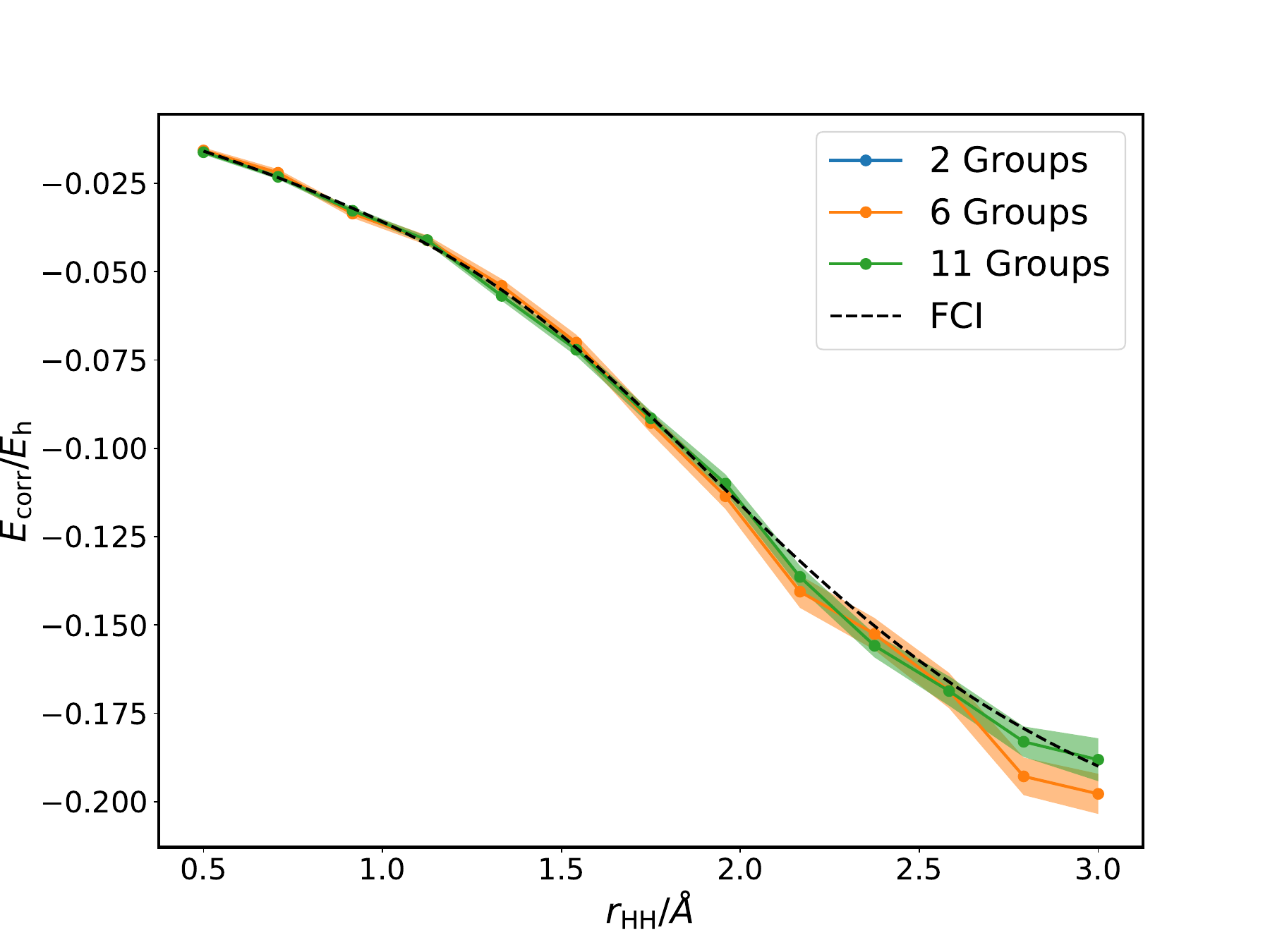}
    \caption{\small \justifying Shift (top) and projected energy (bottom) estimators for the H$_3^+$ correlation energy, obtained using a sampled Hamiltonian with only 2-11 Pauli groups selected at each time-step, a stochastically rounded wavefunction and 1000 shots per residual.} 
    \label{fig:full_stoch_bind}
\end{figure}

{We then consider propagating the full wavefunction with a stochastically sampled Hamiltonian in which only $N_\mathrm{hamil}$ commuting Pauli groups are included at each time-step.} Corresponding values of the estimators for such calculations are given in \Cref{tab:hamil}. Once again, as expected, variance decreases with the number of terms included and we note happily that propagation under the action of the stochastically sampled Hamiltonian does not introduce any additional bias to the estimators. Noise in the shift fully depends on the details of the stochastic trajectory, but in order to minimise the variance of $E_\mathrm{proj}$  we use the full Hamiltonian at each step. As this is  equivalent to a single residual computation, the relative cost of employing the full Hamiltonian at this stage decays with system size. However, the projected energy could also be estimated using the sampled Hamiltonian at each step, decreasing computational overhead at the cost of increased sampling noise.

Considering the data presented above, we are satisfied that shot-based simulation, stochastic wavefunction rounding and Hamiltonian sampling are all compatible with MC-PQE and do not introduce significant biases into the results of a simulation. They do however all increase the noise present, requiring, much like classical QMC methods, a relatively long period of steady-state propagation to converge the desired quantities to the intended accuracy.

There are further ways to reduce the noise in the simulation which do not require increased sampling. One such method is \textit{shift damping},\cite{Booth2009} which entails decreasing the parameter $\zeta$ in \Cref{eq:shift}. This makes the shift less responsive to changes in population, thereby directly decreasing its variance. As the value of the shift affects the continued propagation of the wavefunction, we expect that, once the steady-state has been reached, less variability of the shift will translate to less variability of the wavefunction and therefore one might expect less noise in the projected energy as well. 

As an example, we compare the shift and projected energy in calculations with different numbers of shots and different degrees of shift damping for $H_3^+$. While a 10-fold decrease in  $\zeta$ is very effective in reducing shift noise ($\sigma_\mathrm{damped}/\sigma =$ 18\% (1000 shots), 20\% (100 shots)), we see no effect in the projected energy ($\sigma_\mathrm{damped}/\sigma =$ 101\% (1000 shots), 102\% (100 shots)). See the appendix for example MC-PQE trajectories.

Finally, we combine shot-based simulation with stochastic wavefunction rounding and Hamiltonian sampling. $H_3^+$ binding curves obtained with the fully stochastic algorithm are given in \Cref{fig:full_stoch_bind}. Simulations remain well behaved even with multiple sources of noise. We also note that the systematic increase in error due to stochastic rounding of the wavefunction appears to be partially cancelled out by other noise.  As  noted in the case of pure finite sampling noise, more highly correlated regimes require better sampling to achieve the same level of accuracy.

\begin{table*}
    \centering
    \caption{\small \justifying Maximum absolute error (MAE) and non-parallelity error (NPE) in milliHartree for Monte Carlo PQE in some frozen-core second row hydrides, compared to fully deterministic PQE. These systems have 125 (LiH and HF) and 313 (H$_2$O and BeH$_2$) Hamiltonian groups respectively. Results in all columns have Gaussian noise with $\sigma = 0.001$ added to simulate sampling noise and results are obtained from runs of 2000 imaginary time-steps $\delta \beta = 0.2$. The first set of results (columns 1-4) have no additional stochasticity. The second (5-8) and third (9-12) sets use stochastic wavefunctions and 10 or 20 sampled Hamiltonian terms per step respectively. The error converges towards the pure sampling error with few sampled Hamiltonian groups. For all methods, errors could be further decreased by using longer propagation times.}
    \label{tab:hydrides}
    \begin{tabular}{l||c|c|c|c||c|c|c|c||c|c|c|c}
      \multirow{2}{*}{System} &  \multicolumn{2}{c|}{MAE ($\sigma$)} & \multicolumn{2}{c||}{NPE ($\sigma$)} & \multicolumn{2}{c|}{MAE ($\sigma$, $N_h = 10$)} & \multicolumn{2}{c||}{NPE ($\sigma$, $N_h = 10$)} & \multicolumn{2}{c|}{MAE ($\sigma$, $N_h = 20$)} &\multicolumn{2}{c}{NPE ($\sigma$, $N_h = 20$)}\\
      \cline{2-13}
      & S & E& S & E& S & E& S & E& S & E& S & E\\
      \hline
        LiH & 0.5 & 0.3 & 0.6 & 0.6 & \hspace*{5mm}2\hspace*{5mm} & 1& \hspace*{5mm}4\hspace*{5mm} &1.6&\hspace*{5mm}2\hspace*{5mm} &0.7 &\hspace*{4mm}2.7\hspace*{4mm}&1\\
         BeH$_2$ & 0.6 & 0.9 & 0.5 & 0.9 & 2 & 3 & 5 & 6 &2&3&4&4 \\
         H$_2$O & 6 & 4 & 6& 6& 42 &76 & 53& 72&9&5& 10& 5 \\
         HF& 9 & 4& 14 & 4 & 10& 13& 18 & 20 & 6 & 10 & 6 &14\\
         \hline
    \end{tabular}
    
\end{table*}

We also apply this stochastic PQE algorithm to first row mono- and dihydrides over a range of geometries, with results summarised in \Cref{tab:hydrides}. The cases studied include the dissociation of LiH and HF, the symmetric stretch of H$_2$O at its equilibrium bond-angle and the $C_{2v}$ insertion of Be into the H$_2$ bond to form BeH$_2$.\cite{Purvis1983}
Geometries for HF and H$_2$O match those from \cite{Das2010} and the STO-3G basis set was used for all molecules. When using the same calculation parameters for all systems, it is unsurprising that errors are larger for more complex molecules. However, we note that by stochastically rounding the wavefunction and only sampling 20 Hamiltonian groups for each system, similar errors to the pure finite sampling case are observed. In all cases, the results correspond to relatively short propagation times and errors could be reduced by running the calculations further.

\subsection{Excited States}

As described in Sec II D, the algorithms above can also be applied to excited states by employing the modified Hamiltonian in \Cref{eq:fs_ham}. The results of example folded-spectrum calculations  performed using exact statevector simulations for H$_3^+$ are given in \Cref{fig:fs_h3+}. In order to easily converge to the desired state, the calculations are initialised at the first geometry (shortest $r_\mathrm{HH}$) with different reference determinants and values of $\omega$ close to their energies. 
States are then tracked along the binding curve by using the parameters and energies of the previous point to initialise the wavefunction and $\omega$ value for the current point. This allows smooth tracking of the states over the whole range of geometries considered.

We investigate introducing noise into these simulations, first in the form of pure sampling noise due to shot-based simulation. Results obtained using Gaussian noise with $\sigma = 0.01$ for  H$_3^+$ are shown in the left panel of \Cref{fig:fs_reblock}. We note that, even with relatively realistic finite sampling noise levels, convergence is still very good, with obtained states closely following their FCI  counterparts.

However, not all instances of noise are as benign in this scenario, as can be seen in the right panel of \Cref{fig:fs_reblock}, which shows results for the totally symmetric states of H$_3^+$ obtained with stochastically rounded wavefunctions. In this case, particularly for the highly excited states, the rounding of the wavefunction leads to unphysical behaviour. The third excited state is most strongly affected, at points tending towards converging onto the highest excited state instead. One must note that because previous values of the energies are used as $\omega$ values at subsequent points, once a sufficiently large deviation occurs it is likely to remain or be amplified. This could be avoided by running independent calculations at each geometry, but a new problem of estimating \textit{a priori} a good initial wavefunction and $\omega$ is introduced.

\begin{figure}[h]
    \centering
    \includegraphics[width=0.5\textwidth]{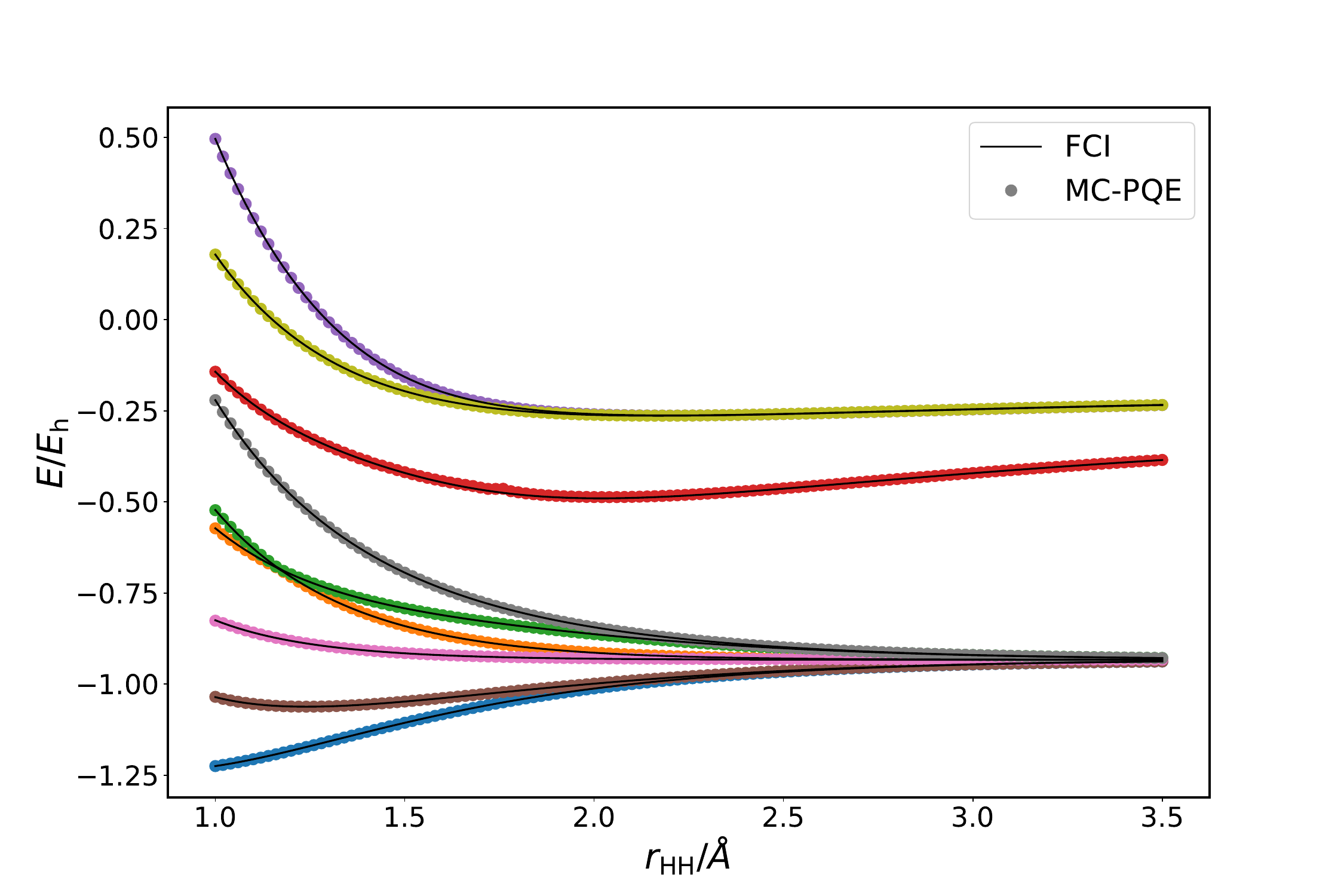}
    \caption{\small \justifying Statevector-simulated folded-spectrum energies for the H$_3^+$ molecule in the STO-3G basis set, using a projective eigenvalue solver. The Slater determinants used as initial references for each excited states are, in increasing energy order of the resulting state: $\ket{01}$, $\ket{03}$, $\ket{12}$, $\ket{23}$, $\ket{05}$, $\ket{14}$, $\ket{25}$, $\ket{34}$, $\ket{45}$, following the same conventions defined for \Cref{eq:trial_wfn}.}
    \label{fig:fs_h3+}
\end{figure}

\begin{figure*}
    \centering
    \includegraphics[width=\textwidth,trim = 3.5cm 0cm 5cm 2.1cm, clip]{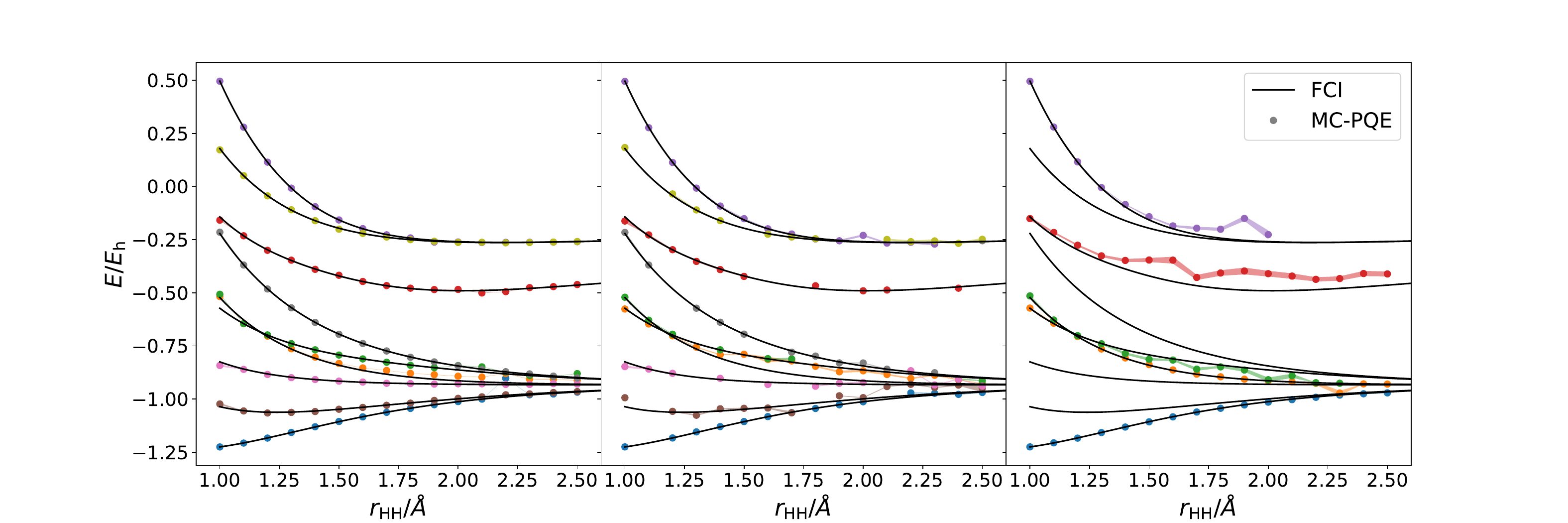}
    \caption{\small \justifying  Folded-spectrum energies for the H$_3^+$ molecule in the STO-3G basis set, using a projective eigensolver and (left) Gaussian noise with $\sigma = 0.01$, (middle) Gaussian noise with $\sigma = 0.01$ and 10 Hamiltonian groups sampled at each step, (right) Gaussian noise with $\sigma = 0.01$ and a stochastically rounded wavefunction.}
    \label{fig:fs_reblock}
\end{figure*}

The main limitation to using the folded spectrum method is the increase in terms in the squared Hamiltonian compared to the original operator. {However, the Monte Carlo stochastic sampling of the Hamiltonian described in \cref{sec:hamil} provides an effective means of sampling the folded spectrum operator. We find (see middle panel of \Cref{fig:fs_reblock}) that the same number of sampled Hamiltonian terms can be used per step as in standard ground state calculations to generate good quality excited state energies. This significantly expands on the range of systems for which folded spectrum methods are tractable.}

\section{Conclusion}
In this paper, we present a Quantum Monte Carlo-informed approach to projective quantum eigensolver algorithms. We begin by noting that rather than attempting to obtain the true ground state of a system to within the desired accuracy directly, once propagation has converged, we can obtain highly accurate estimates by classically averaging over samples obtained at different times over the course of imaginary time propagation. This allows us to use low shot counts without compromising final accuracy, leading to a two orders of magnitude reduction in the number of samples relative to comparable conventional VQE calculations. Notably, this reduction is due to an effective increase in the target variance of the measured observable, and therefore orthogonal to potential improvements due to more effective utilisation of the measurement information itself. Therefore, this method should be usable in conjunction with efficient measurement techniques such as those based on classical shadows to great effect.

We go further to decrease the quantum computational overhead, through stochastic rounding of the wavefunction and probabilistic sampling of the Hamiltonian. These procedures lead to shallower circuits and fewer shots required per step, but also generate noisier projectors.  However, we observe that the Monte Carlo energy estimators employed are generally resilient to this additional noise, generating comparable accuracy to the deterministic projector, provided the resulting truncation at each step is not too drastic. In practice, for second row hydrides we observe that as few as 10\% of the Hamiltonian terms are required.

Finally, we expand this method to excited state calculations, by employing the folded-spectrum approach, in which the Hamiltonian is replaced by a variance operator.  The squared operator has significantly more terms than the underlying Hamiltonian, making this method traditionally more expensive than conventional eigensolvers. However, by sampling the operator we find that we can reduce the cost to that of an equivalent ground state calculation.

The fundamental reason these stochastic approximations and sampling methods work is that, given a wavefunction that is a steady-state of some operator, they induce fluctuations about this wavefunction that may be effectively averaged over. Therefore, their applicability is not limited to the projective quantum eigensolver and we believe they may be helpful in any quantum algorithm which aims to converge to a some well-defined final wavefunction.

\section*{Acknowledgements}
The author thanks Alex Thom and Lila Cadi-Tazi for useful discussions and gratefully acknowledges Peterhouse, University of Cambridge for supporting this work through a Research Fellowship.

\bibliography{qc_qmc}
\appendix
\section{Example MC-PQE trajectories.}

In this appendix we present the MC-PQE trajectories used to generate the data in \Cref{tab:hamil}, as well as to estimate the influence of shift damping on the overall noise of a simulation. 

\Cref{fig:hamil} shows $S$ and $E_\mathrm{proj}$ over the course of a MC-PQE simulation in which the residuals are computed deterministically, but the Hamiltonian is sampled, for H$_3^+$. We note that as the number of terms in the sampled Hamiltonian increases, the deviations of the instantaneous estimators from the true energy decay, as expected. This corresponds to decreasing variance of the overall energy estimators.

\Cref{fig:full_stoch} shows the same data for a fully stochastic MC-PQE calculation, with 1000 shots per measurement. Similar trends are observed with increasing number of Hamiltonian groups.

\Cref{fig:shift_damping} shows the effect of decreasing the damping factor $\zeta$ in \Cref{eq:shift} on the fluctuations of the shift and projected energy estimators. While a 10$\times$ decrease in $\zeta$ reduces the variance of the shift estimator by approximately a factor of 5, it has no effect on $E_\mathrm{proj}$, as discussed in more detail in the main text.

\onecolumngrid

\begin{figure}
    \centering
    \includegraphics[width=0.45 \textwidth, trim = 0.55cm 0.5cm 3.8cm 2cm, clip]{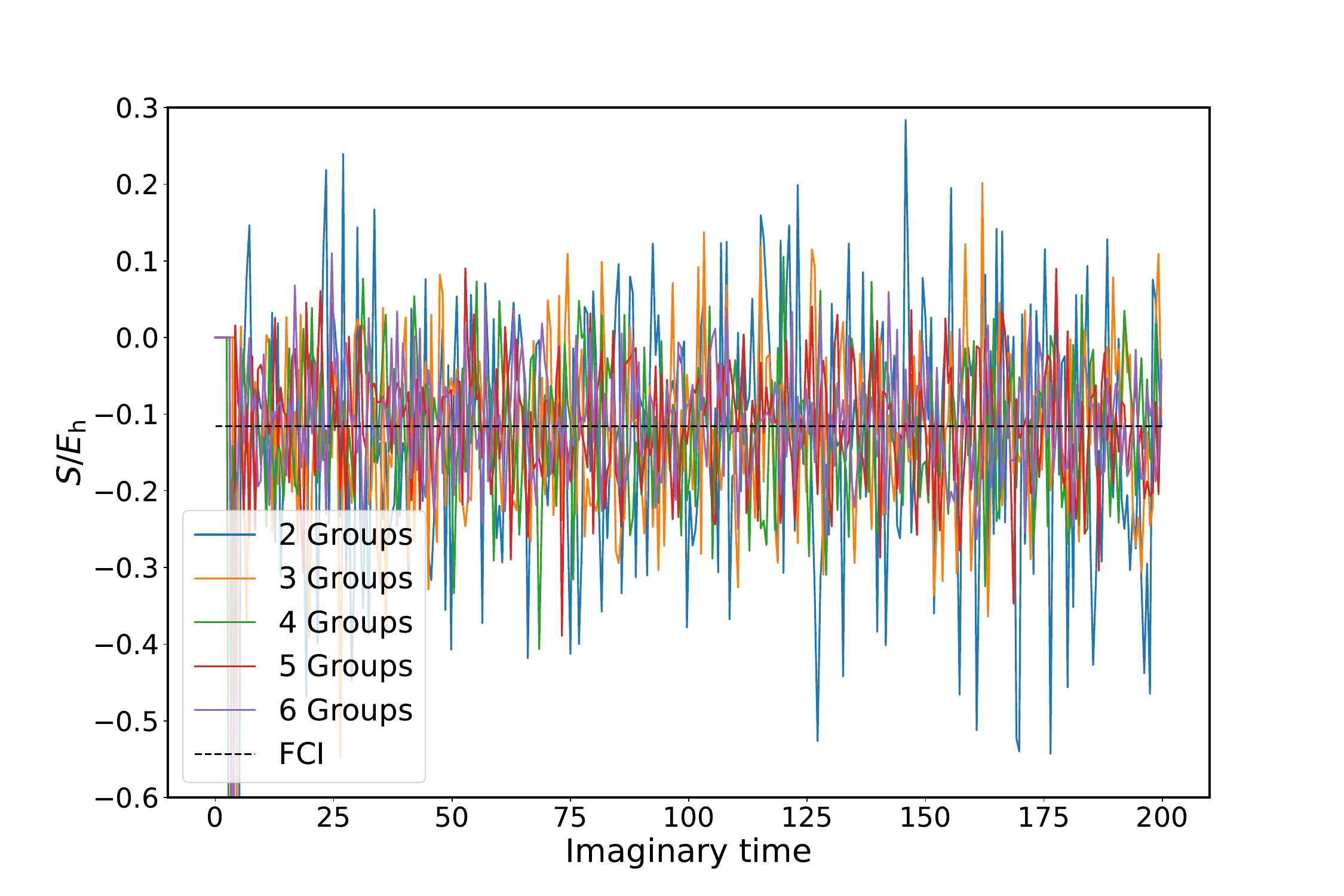}
    \includegraphics[width=0.45 \textwidth, trim = 0.55cm 0.5cm 3.8cm 2cm, clip]{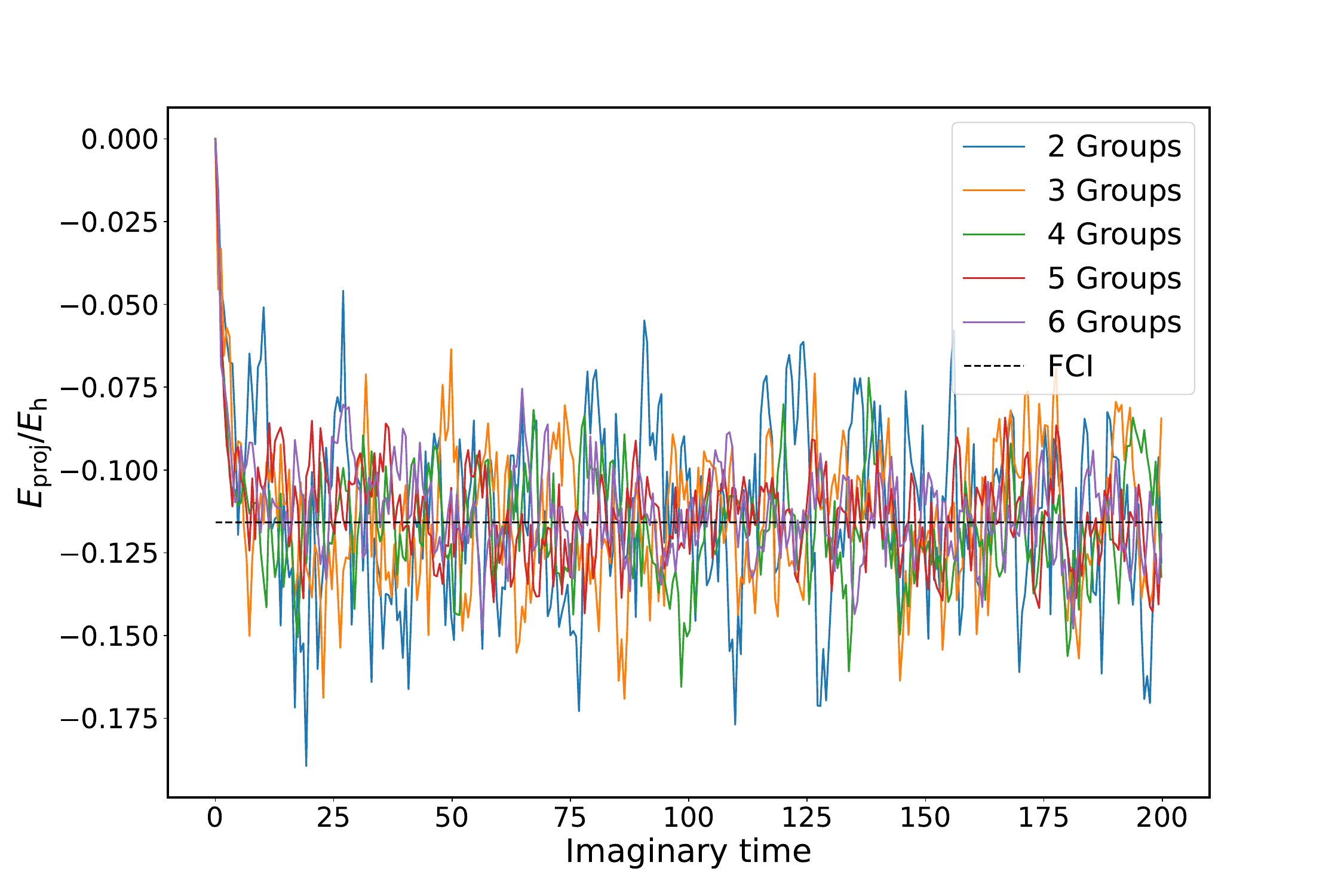}
    \caption{\small \justifying Shift (left) and projected energy (right) as a function of imaginary time obtained by imaginary-time propagation using a stochastically sampled Hamiltonian with only 2-6 Pauli groups selected at each time-step for H$_3^+$ at $r = 2.0$ \r{A}. The full H$_3^+$ Hamiltonian has 62 Pauli terms, split here into 25 commuting groups.}
    \label{fig:hamil}
\end{figure}

\begin{figure}
    \centering
    \includegraphics[width=0.45 \textwidth, trim = 0.65cm 0.7cm 3.8cm 2cm, clip]{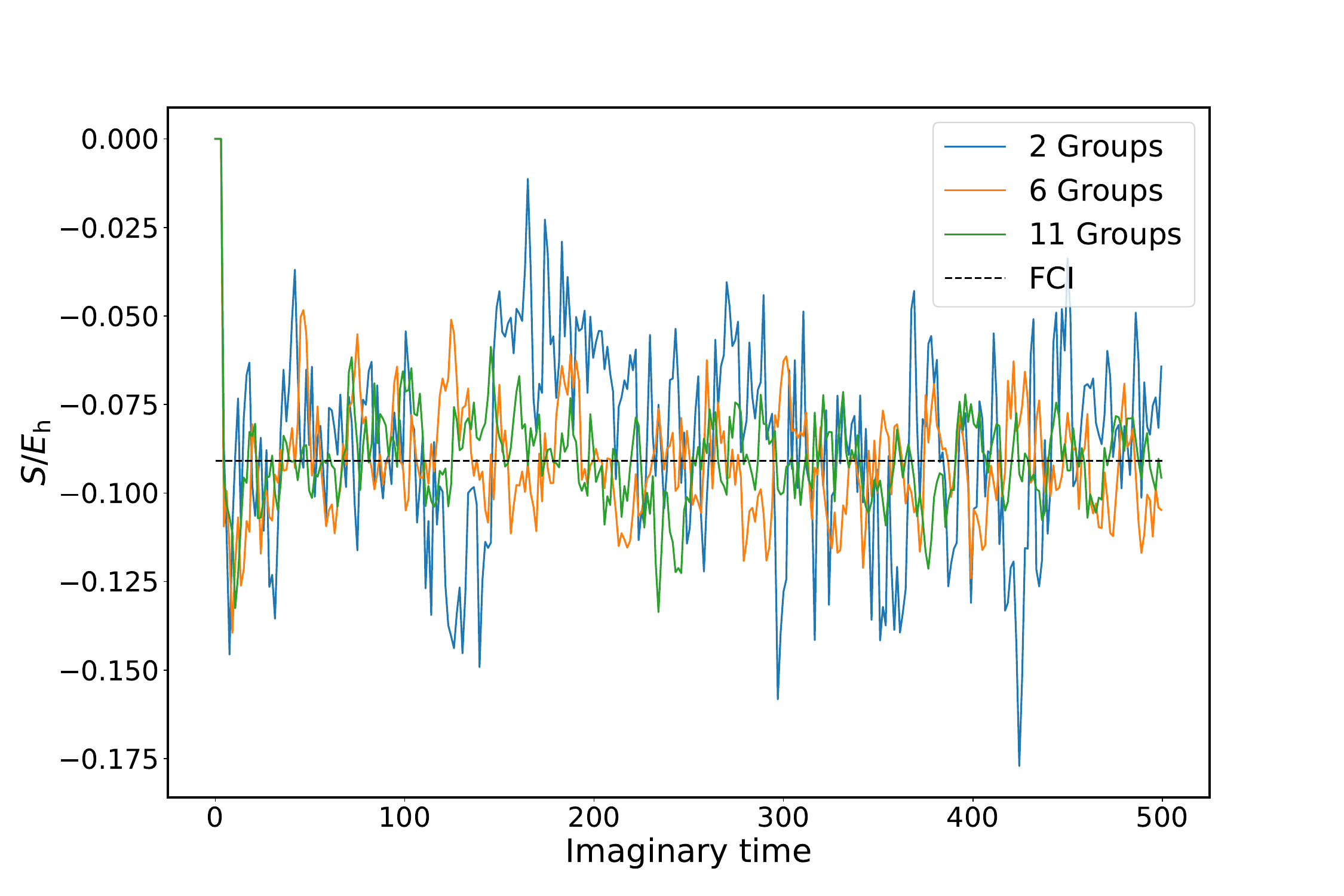}
    \includegraphics[width=0.45 \textwidth, trim = 0.65cm 0.7cm 3.8cm 2cm, clip]{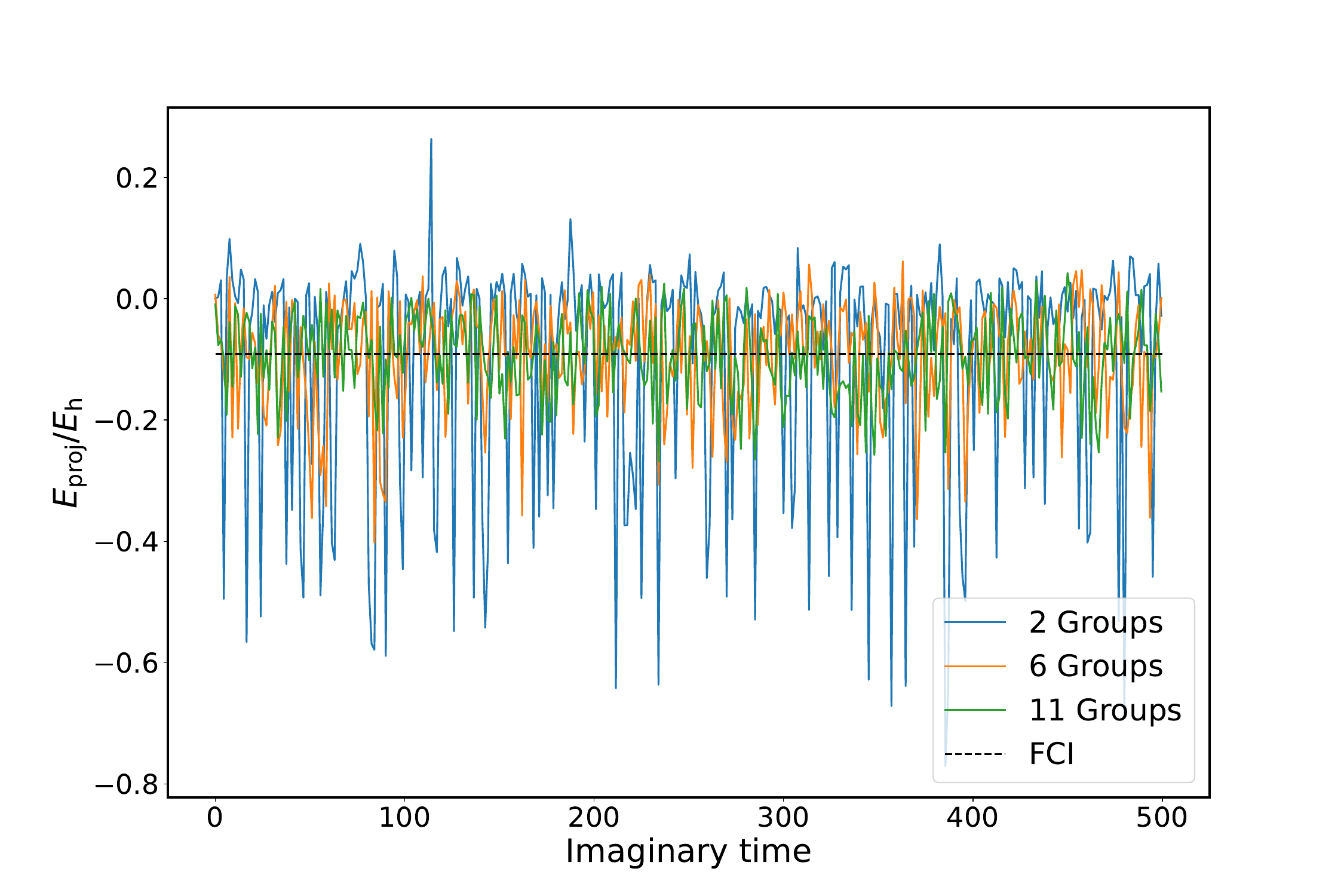}
    \caption{\small \justifying Shift (left) and projected energy (right) as a function of imaginary time obtained by imaginary-time propagation using a stochastically sampled Hamiltonian with only 2-11 Pauli groups selected at each time-step, a stochastically rounded wavefunction and 1000 shots.} 
    \label{fig:full_stoch}
\end{figure}

\begin{figure}
    \centering
    \includegraphics[width=0.49 \textwidth, trim = 1cm 2cm 3.6cm 2.5cm, clip]{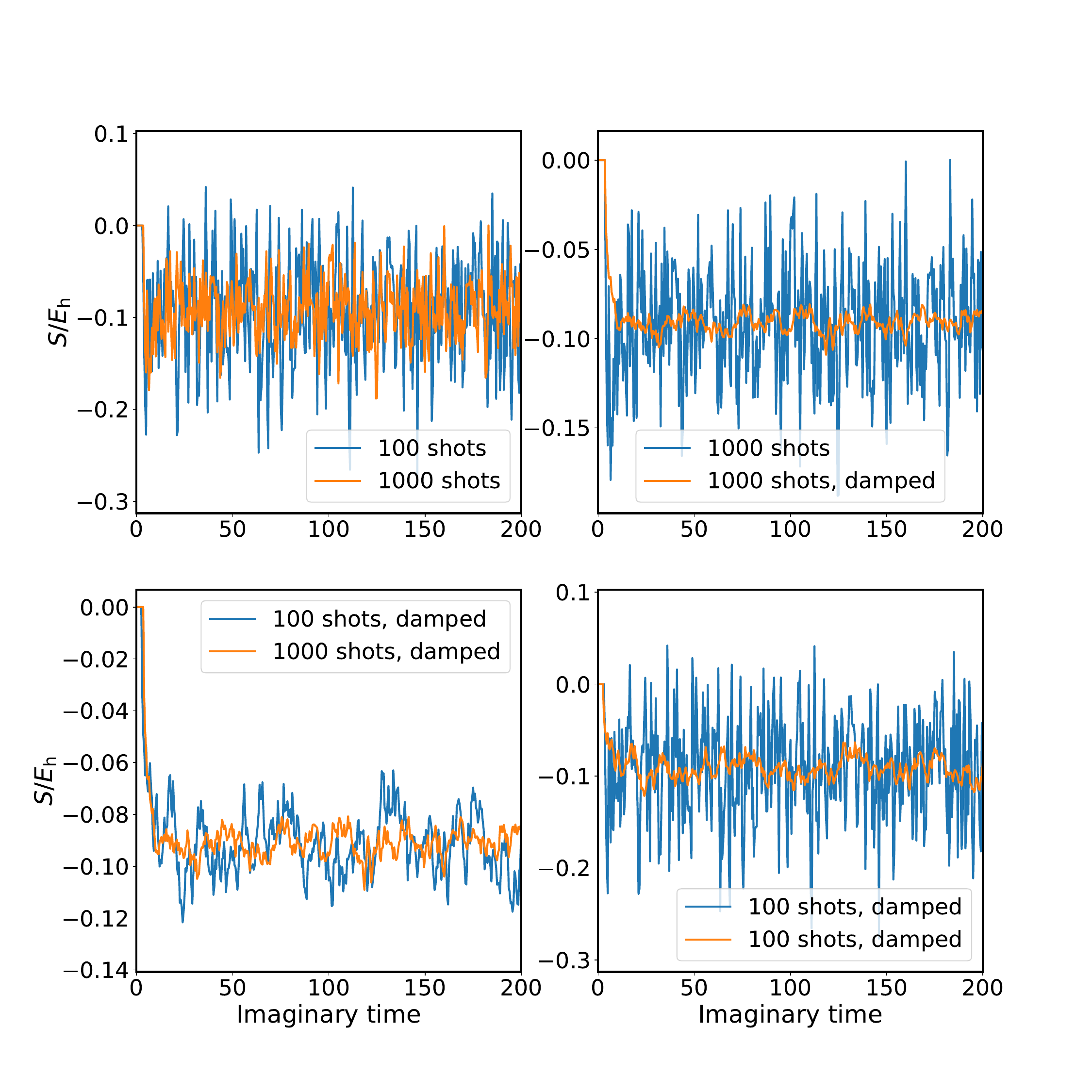}
    \includegraphics[width=0.49 \textwidth, trim = 1cm 2cm 3.6cm 2.5cm, clip]{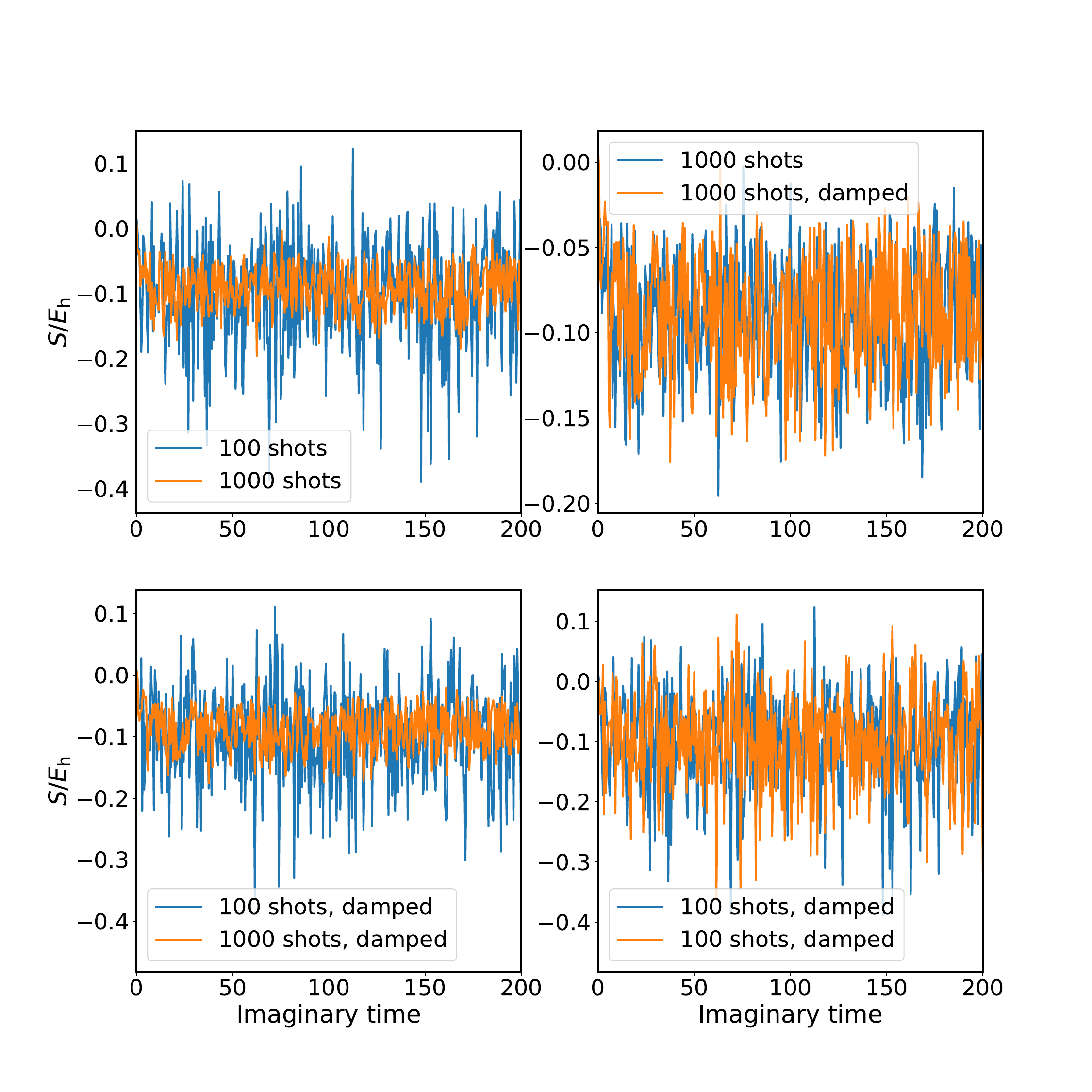}
    \caption{\small \justifying Shift (4 left panels) and $E_\mathrm{proj}$ (4 right panels) in a simulation of H$_3^+$ at $r = 1.75$, using 100 or 1000 shots per residual estimation and shift damping parameters $\zeta = 1$ in the undamped case and $\zeta=0.1$ in the damped case. Shift damping clearly reduces shift fluctuations, while leaving those in the projected energy unchanged.}
    \label{fig:shift_damping}
\end{figure}

\end{document}